\documentclass[twocolumn]{aastex7}

\newcommand\lya{Ly$\alpha$}
\newcommand\delLAE{$\delta_{\rm LAE}$}

\usepackage{lipsum}

\begin{document}

\title{Effect of local environment on Ly$\alpha$ line profile in DESI/ODIN LAEs}

\author[0000-0001-9308-0449]{Ana Sof{\'i}a M. Uzsoy}
\affiliation{Center for Astrophysics $|$ Harvard \& Smithsonian, 60 Garden St., Cambridge, MA 02138, USA}
\email[show]{ana\_sofia.uzsoy@cfa.harvard.edu}

\author[0000-0002-4928-4003]{Arjun~Dey}
\affiliation{NSF NOIRLab, 950 N. Cherry Ave., Tucson, AZ 85719, USA}
\email{}

\author[0000-0001-5999-7923]{Anand~Raichoor}
\affiliation{Lawrence Berkeley National Laboratory, 1 Cyclotron Road, Berkeley, CA 94720, USA}
\email{}

\author[0000-0003-2808-275X]{Douglas P. Finkbeiner}
\affiliation{Center for Astrophysics $|$ Harvard \& Smithsonian, 60 Garden St., Cambridge, MA 02138, USA}
\affiliation{Department of Physics, Harvard University, 17 Oxford St., Cambridge, MA 02138, USA}
\email{}

\author[0000-0003-3004-9596]{Vandana Ramakrishnan}
\affiliation{Department of Physics and Astronomy, Purdue University, 525 Northwestern Ave., West Lafayette, IN 47906, USA}
\email{}

\author[0000-0003-3004-9596]{Kyoung-Soo Lee}
\affiliation{Department of Physics and Astronomy, Purdue University, 525 Northwestern Ave., West Lafayette, IN 47906, USA}
\email{}

\author[0000-0003-1530-8713]{Eric Gawiser}
\affiliation{Department of Physics and Astronomy, Rutgers, the State University of New Jersey, Piscataway, NJ 08854, USA}
\email{}

\author{Jessica~Nicole~Aguilar}
\affiliation{Lawrence Berkeley National Laboratory, 1 Cyclotron Road, Berkeley, CA 94720, USA}
\email{}

\author[0000-0001-6098-7247]{Steven~Ahlen}
\affiliation{Physics Dept., Boston University, 590 Commonwealth Avenue, Boston, MA 02215, USA}
\email{}

\author[0000-0003-2923-1585]{Abhijeet~Anand}
\affiliation{Lawrence Berkeley National Laboratory, 1 Cyclotron Road, Berkeley, CA 94720, USA}
\email{}

\author[0000-0001-9712-0006]{Davide~Bianchi}
\affiliation{Dipartimento di Fisica ``Aldo Pontremoli'', Universit\`a degli Studi di Milano, Via Celoria 16, I-20133 Milano, Italy}
\affiliation{INAF-Osservatorio Astronomico di Brera, Via Brera 28, 20122 Milano, Italy}
\email{}

\author{David~Brooks}
\affiliation{Department of Physics \& Astronomy, University College London, Gower Street, London, WC1E 6BT, UK}
\email{}

\author{Todd~Claybaugh}
\affiliation{Lawrence Berkeley National Laboratory, 1 Cyclotron Road, Berkeley, CA 94720, USA}
\email{}

\author[0000-0002-1769-1640]{Axel~de la Macorra}
\affiliation{Instituto de F\'{\i}sica, Universidad Nacional Aut\'{o}noma de M\'{e}xico,  Circuito de la Investigaci\'{o}n Cient\'{\i}fica, Ciudad Universitaria, Cd. de M\'{e}xico  C.~P.~04510,  M\'{e}xico}
\email{}

\author{Peter~Doel}
\affiliation{Department of Physics \& Astronomy, University College London, Gower Street, London, WC1E 6BT, UK}
\email{}

\author[0000-0003-4992-7854]{Simone~Ferraro}
\affiliation{Lawrence Berkeley National Laboratory, 1 Cyclotron Road, Berkeley, CA 94720, USA}
\affiliation{University of California, Berkeley, 110 Sproul Hall \#5800 Berkeley, CA 94720, USA}
\email{}

\author[0000-0002-9811-2443]{Nicole~M.~Firestone}
\affiliation{Department of Physics and Astronomy, Rutgers, the State University of New Jersey, Piscataway, NJ 08854, USA}
\email{}

\author[0000-0002-3033-7312]{Andreu~Font-Ribera}
\affiliation{Institut de F\'{i}sica d'Altes Energies (IFAE), The Barcelona Institute of Science and Technology, Edifici Cn, Campus UAB, 08193, Bellaterra (Barcelona), Spain}
\email{}

\author[0000-0002-2890-3725]{Jaime~E.~Forero-Romero}
\affiliation{Departamento de F\'isica, Universidad de los Andes, Cra. 1 No. 18A-10, Edificio Ip, CP 111711, Bogot\'a, Colombia}
\affiliation{Observatorio Astron\'omico, Universidad de los Andes, Cra. 1 No. 18A-10, Edificio H, CP 111711 Bogot\'a, Colombia}
\email{}

\author{Enrique~Gazta\~{n}aga}
\affiliation{Institut d'Estudis Espacials de Catalunya (IEEC), c/ Esteve Terradas 1, Edifici RDIT, Campus PMT-UPC, 08860 Castelldefels, Spain}
\affiliation{Institute of Cosmology and Gravitation, University of Portsmouth, Dennis Sciama Building, Portsmouth, PO1 3FX, UK}
\affiliation{Institute of Space Sciences, ICE-CSIC, Campus UAB, Carrer de Can Magrans s/n, 08913 Bellaterra, Barcelona, Spain}
\email{}

\author[0000-0002-4902-0075]{Lucia~Guaita}
\affiliation{Universidad Andres Bello, Facultad de Ciencias Exactas, Departamento de Fisica y Astronomia, Instituto de Astrofisica, Fernandez Concha 700, Las Condes, Santiago RM, Chile}
\affiliation{Millennium Nucleus for Galaxies (MINGAL)}
\email{}

\author{Gaston~Gutierrez}
\affiliation{Fermi National Accelerator Laboratory, PO Box 500, Batavia, IL 60510, USA}
\email{}

\author[0000-0002-9136-9609]{Hiram~K.~Herrera-Alcantar}
\affiliation{Institut d'Astrophysique de Paris. 98 bis boulevard Arago. 75014 Paris, France}
\affiliation{IRFU, CEA, Universit\'{e} Paris-Saclay, F-91191 Gif-sur-Yvette, France}
\email{}

\author[0000-0002-6550-2023]{Klaus~Honscheid}
\affiliation{Center for Cosmology and AstroParticle Physics, The Ohio State University, 191 West Woodruff Avenue, Columbus, OH 43210, USA}
\affiliation{Department of Physics, The Ohio State University, 191 West Woodruff Avenue, Columbus, OH 43210, USA}
\affiliation{The Ohio State University, Columbus, 43210 OH, USA}
\email{}

\author[0000-0002-6024-466X]{Ho~Seong~Hwang}
\affiliation{Astronomy Program, Department of Physics and Astronomy,
Seoul National University, 1 Gwanak-ro, Gwanak-gu, Seoul 08826, Republic
of Korea}
\affiliation{SNU Astronomy Research Center, Seoul National University, 1
Gwanak-ro, Gwanak-gu, Seoul 08826, Republic of Korea}
\email{}

\author[0000-0002-6024-466X]{Mustapha~Ishak}
\affiliation{Department of Physics, The University of Texas at Dallas, 800 W. Campbell Rd., Richardson, TX 75080, USA}
\email{}

\author[0000-0003-0201-5241]{Dick~Joyce}
\affiliation{NSF NOIRLab, 950 N. Cherry Ave., Tucson, AZ 85719, USA}
\email{}

\author[0000-0002-8828-5463]{David~Kirkby}
\affiliation{Department of Physics and Astronomy, University of California, Irvine, 92697, USA}
\email{}

\author[0000-0003-3510-7134]{Theodore~Kisner}
\affiliation{Lawrence Berkeley National Laboratory, 1 Cyclotron Road, Berkeley, CA 94720, USA}
\email{}

\author[0000-0001-6356-7424]{Anthony~Kremin}
\affiliation{Lawrence Berkeley National Laboratory, 1 Cyclotron Road, Berkeley, CA 94720, USA}
\email{}

\author{Ofer~Lahav}
\affiliation{Department of Physics \& Astronomy, University College London, Gower Street, London, WC1E 6BT, UK}
\email{}

\author[0000-0002-6731-9329]{Claire~Lamman}
\affiliation{The Ohio State University, Columbus, 43210 OH, USA}
\email{}

\author[0000-0003-1838-8528]{Martin~Landriau}
\affiliation{Lawrence Berkeley National Laboratory, 1 Cyclotron Road, Berkeley, CA 94720, USA}
\email{}

\author[0000-0001-7178-8868]{Laurent~Le~Guillou}
\affiliation{Sorbonne Universit\'{e}, CNRS/IN2P3, Laboratoire de Physique Nucl\'{e}aire et de Hautes Energies (LPNHE), FR-75005 Paris, France}
\email{}

\author[0000-0003-4962-8934]{Marc~Manera}
\affiliation{Departament de F\'{i}sica, Serra H\'{u}nter, Universitat Aut\`{o}noma de Barcelona, 08193 Bellaterra (Barcelona), Spain}
\affiliation{Institut de F\'{i}sica d’Altes Energies (IFAE), The Barcelona Institute of Science and Technology, Edifici Cn, Campus UAB, 08193, Bellaterra (Barcelona), Spain}
\email{}

\author{Ramon~Miquel}
\affiliation{Instituci\'{o} Catalana de Recerca i Estudis Avan\c{c}ats, Passeig de Llu\'{\i}s Companys, 23, 08010 Barcelona, Spain}
\affiliation{Institut de F\'{i}sica d'Altes Energies (IFAE), The Barcelona Institute of Science and Technology, Edifici Cn, Campus UAB, 08193, Bellaterra (Barcelona), Spain}
\email{}

\author[0000-0002-2733-4559]{John~Moustakas}
\affiliation{Department of Physics and Astronomy, Siena College, 515 Loudon Road, Loudonville, NY 12211, USA}
\email{}

\author{Andrea~Mu\~{n}oz-Guti\'errez}
\affiliation{Instituto de F\'{\i}sica, Universidad Nacional Aut\'{o}noma de M\'{e}xico,  Circuito de la Investigaci\'{o}n Cient\'{\i}fica, Ciudad Universitaria, Cd. de M\'{e}xico  C.~P.~04510,  M\'{e}xico}
\email{}

\author[0000-0001-9070-3102]{Seshadri~Nadathur}
\affiliation{Institute of Cosmology and Gravitation, University of Portsmouth, Dennis Sciama Building, Portsmouth, PO1 3FX, UK}
\email{}

\author[0000-0003-3188-784X]{Nathalie~Palanque-Delabrouille}
\affiliation{IRFU, CEA, Universit\'{e} Paris-Saclay, F-91191 Gif-sur-Yvette, France}
\affiliation{Lawrence Berkeley National Laboratory, 1 Cyclotron Road, Berkeley, CA 94720, USA}
\email{}

\author[0000-0002-0644-5727]{Will~Percival}
\affiliation{Department of Physics and Astronomy, University of Waterloo, 200 University Ave W, Waterloo, ON N2L 3G1, Canada}
\affiliation{Perimeter Institute for Theoretical Physics, 31 Caroline St. North, Waterloo, ON N2L 2Y5, Canada}
\affiliation{Waterloo Centre for Astrophysics, University of Waterloo, 200 University Ave W, Waterloo, ON N2L 3G1, Canada}
\email{}

\author{Claire~Poppett}
\affiliation{Lawrence Berkeley National Laboratory, 1 Cyclotron Road, Berkeley, CA 94720, USA}
\affiliation{Space Sciences Laboratory, University of California, Berkeley, 7 Gauss Way, Berkeley, CA  94720, USA}
\affiliation{University of California, Berkeley, 110 Sproul Hall \#5800 Berkeley, CA 94720, USA}
\email{}

\author[0000-0001-7145-8674]{Francisco~Prada}
\affiliation{Instituto de Astrof\'{i}sica de Andaluc\'{i}a (CSIC), Glorieta de la Astronom\'{i}a, s/n, E-18008 Granada, Spain}
\email{}

\author[0000-0001-6979-0125]{Ignasi~P\'erez-R\`afols}
\affiliation{Departament de F\'isica, EEBE, Universitat Polit\`ecnica de Catalunya, c/Eduard Maristany 10, 08930 Barcelona, Spain}
\email{}

\author{Graziano~Rossi}
\affiliation{Department of Physics and Astronomy, Sejong University, 209 Neungdong-ro, Gwangjin-gu, Seoul 05006, Republic of Korea}
\email{}

\author[0000-0002-9646-8198]{Eusebio~Sanchez}
\affiliation{CIEMAT, Avenida Complutense 40, E-28040 Madrid, Spain}
\email{}

\author{David~Schlegel}
\affiliation{Lawrence Berkeley National Laboratory, 1 Cyclotron Road, Berkeley, CA 94720, USA}
\email{}

\author{Michael~Schubnell}
\affiliation{Department of Physics, University of Michigan, 450 Church Street, Ann Arbor, MI 48109, USA}
\affiliation{University of Michigan, 500 S. State Street, Ann Arbor, MI 48109, USA}
\email{}

\author[0000-0002-6588-3508]{Hee-Jong~Seo}
\affiliation{Department of Physics \& Astronomy, Ohio University, 139 University Terrace, Athens, OH 45701, USA}
\email{}

\author[0000-0002-3461-0320]{Joseph~Harry~Silber}
\affiliation{Lawrence Berkeley National Laboratory, 1 Cyclotron Road, Berkeley, CA 94720, USA}
\email{}

\author[0000-0002-4362-4070]{Hyunmi~Song}
\affiliation{Department of Astronomy and Space Science and Institute for Sciences of the Universe, Chungnam National University, Daejeon 34134, Republic of Korea}
\email{}

\author{David~Sprayberry}
\affiliation{NSF NOIRLab, 950 N. Cherry Ave., Tucson, AZ 85719, USA}
\email{}

\author[0000-0003-1704-0781]{Gregory~Tarl\'{e}}
\affiliation{University of Michigan, 500 S. State Street, Ann Arbor, MI 48109, USA}
\email{}

\author{Benjamin~Alan~Weaver}
\affiliation{NSF NOIRLab, 950 N. Cherry Ave., Tucson, AZ 85719, USA}
\email{}

\author[0000-0002-6684-3997]{Hu~Zou}
\affiliation{National Astronomical Observatories, Chinese Academy of Sciences, A20 Datun Road, Chaoyang District, Beijing, 100101, P.~R.~China}
\email{}

\begin{abstract}

Lyman-Alpha Emitters (LAEs) are star-forming galaxies with significant \lya{} emission and are often used as tracers of large-scale structure at high redshift. We explore the relationship between the \lya{} line profile and environmental density with spectroscopy from the Dark Energy Spectroscopic Instrument (DESI) of LAEs selected with narrow-band photometry through the One-hundred-deg$^2$ DECam Imaging in Narrowbands (ODIN) survey. We use LAE surface density maps in the N419 (z $\sim$ 2.45) and N501 (z $\sim$ 3.12) narrow bands to probe the relationship between local environmental density and the \lya{} line profile. In both narrow bands, we stack the LAE spectra in bins of environmental density and inside and outside of protocluster regions. The N501 data shows $\sim$15\% higher \lya{} line luminosity for galaxies in protoclusters, suggesting increased star formation in these regions. However, the line luminosity is not appreciably greater in protocluster galaxies in the N419 band, suggesting a potential redshift evolution of this effect. The shape of the line profile itself does not vary with environmental density, suggesting that line shape changes are caused by local effects independent of a galaxy's environment. These data suggest a potential relationship between LAE local environmental density, ionized gas distribution, and \lya{} line luminosity.
\end{abstract}



\section{Introduction} 

Lyman-Alpha Emitters (LAEs) are typically star-forming, metal-poor galaxies with significant \lya{} emission that are important cosmological probes \citep{partridge_are_1967, hayes_lyman_2015,  ouchi_observations_2020}. The LAE population evolves with redshift to closely track the onset of the epoch of reionization at $z = 6-7$ \citep{ouchi_statistics_2010}. They also serve as high-redshift tracers of large-scale structure \citep{huang_evaluating_2022, white_clustering_2024, ebina2025}.

Variations in the shape of the \lya{} line of an LAE can provide important information about its physical context at numerous scales. Clumpiness of the interstellar medium \citep[ISM;][]{Duval+14, verhamme_lyman-_2012} and the geometry of the surrounding gas \citep{almadamonter_crossing_2024, AlmadaMonter+25} can cause asymmetric line shapes and the presence of multiple peaks. The hydrogen column density can denote shifts in the location of the \lya{} line peak relative to the systemic redshift, and the expansion velocity and resonant scattering can affect the blue-to-red peak ratio \citep{verhamme_3d_2006, khoraminezhad_simulating_2025, Smith+25}. The same galaxy can produce significantly different emission (and thus different line profile shapes) when observed along varying lines of sight, with \lya{} photons from edge-on galaxies having to traverse a much thicker medium than face-on galaxies \citep{zheng_radiative_2010,verhamme_lyman-_2012, blaizot_simulating_2023}. At increasing redshifts, the blue side of the \lya{} line
is attenuated by the IGM, becoming steeper with more suppression of any blue peak \citep{hayes_spectral_2021}.

Some physical properties of LAEs can be determined empirically from their observed line profiles. The equivalent width (EW) of the \lya{} line negatively correlates with galaxy mass and star-formation rate \citep[SFR;][]{oyarzun_comprehensive_2017, mccarron_stellar_2022}. An LAE's systemic velocity offset can be determined by the width of its emission line \citep{verhamme_recovering_2018}. Comparison of the \lya{} line with metal absorption features allows for identification and characterization of outflows \citep{nianias+25}. 


Protoclusters are high-redshift, star-forming progenitors of low-redshift, quiescent galaxy clusters \citep{alberts_noble_22}. Previous work has shown that protoclusters have significantly higher stellar masses and SFRs than can be detected from their individual galaxies \citep{Popescu+23}. Several protocluster regions of LAEs have been detected \citep[e.g.][]{steidel+00, venemans+07, oteo+18, higuchi+19}, including at least one with increased luminosity that suggests elevated SFR
\citep{lee_discovery_2014, dey_spectroscopic_2016}. Similarly, the \lya{} luminosity function is elevated inside of LAE protocluster regions \citep{nagaraj_odin_2025}.


The One-hundred-deg$^2$ DECam Imaging in Narrowbands (ODIN) survey has identified tens of thousands of LAEs with narrow-band (NB) photometry around redshifts 2.45, 3.12, and 4.55 \citep{lee_one-hundred-deg2_2024, firestone_odin_2024}. This dataset alone has enabled several important advances in LAE science, including studies of their star formation histories \citep{firestone_odin_2025}, clustering analyses \citep{white_clustering_2024, herrera_odin_2025}, identifications of \lya{} ``blobs" and protoclusters \citep{ramakrishnan_odin_2023,
 ramakrishnan+24, ramakrishnan_odin_2025, ramakrishnan_complexA}, and detailed analyses of their luminosity functions \citep{nagaraj_odin_2025}.

The use of narrow-band photometry for selection leads to a sample of LAEs that are essentially in a two-dimensional redshift ``slice", making this dataset particularly well-suited for studies of environmental dependence. This also allows for the separation of local effects from a galaxy's immediate neighbors from global effects, such as the attenuation from the intergalactic medium, that occur on cosmological scales. The combination of NB selection of LAEs with spectroscopy from the Dark Energy Spectroscopic Instrument (DESI) allows for detailed characterization of shape of the \lya{} line profile. In anticipation of large-scale surveys such as DESI-2, which aims to observe over 1.5 million LAEs \citep{schlegel_spectroscopic_2022}, it is important to understand the underlying physics of LAEs to solidify their potential as cosmological probes.

In this work, we investigate the relationship between the local environmental density and the shape of the \lya{} line profile using DESI spectroscopy of ODIN LAEs. An outline of this paper is as follows: in Section~\ref{sec:data} we describe the DESI spectroscopy of the ODIN LAEs and their associated redshifts and local density estimates; in Section~\ref{sec:results} we show that LAEs in overdense regions have higher \lya{} line luminosities at $z \approx 3.12$ but no significant difference at $z \approx 2.45$, and further discuss potential underlying physical processes in Section~\ref{sec:discussion} before concluding in Section~\ref{sec:conclusion}.

\begin{figure}[ht!]
\epsscale{1.2}
\plotone{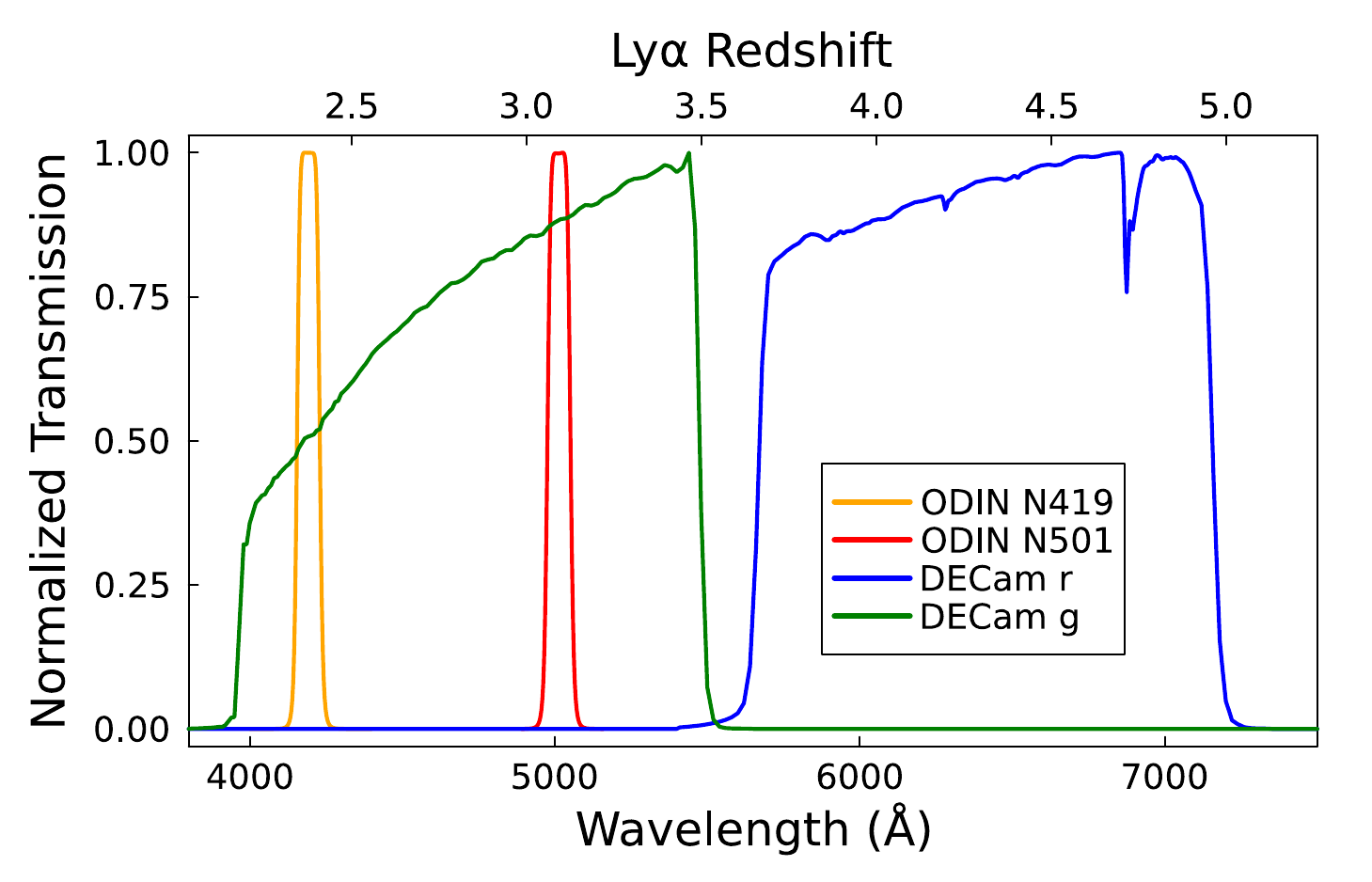}
\caption{Transmission curves for narrow- and broad-band filters used in this analysis. Curves are normalized to a maximum transmission of 1.}
\label{fig:filters}
\end{figure}

\section{Data}
\label{sec:data}

\subsection{LAE Candidate Selection and Spectroscopy}

Here, we use a sample of LAEs selected from the ODIN survey \citep[see][and references therein]{lee_one-hundred-deg2_2024} which have been spectroscopically confirmed using DESI \citep{DESI_Instrument_Overview} on the Nicholas Mayall 4-meter Telescope at the Kitt Peak National Observatory \citep{dey_DESI_ODIN, pinarski}. ODIN provides deep imaging (5$\sigma$ depth of $\approx25-26$AB mag) in three narrow-band filters N419 ($\lambda_{\rm cen}/\Delta\lambda$=4193\AA/74.6\AA), N501 (5014\AA/75.6\AA) and N673 (6750\AA/100\AA) in eight fields using the Dark Energy Camera \citep[DECam;][]{DECam_Instrument} on the Victor Blanco 4-meter Telescope at the Cerro-Tololo Inter-American Observatory \citep[see][for details]{lee_one-hundred-deg2_2024}. Details of the target selection and spectroscopic follow-up of ODIN sources are presented in \cite{lee_one-hundred-deg2_2024} and \cite{dey_DESI_ODIN}, but we summarize the salient points here. 

We constructed photometric catalogs \citep[using Tractor;][]{Tractor2016} of sources detected in the stacked ODIN imaging  in the 10~deg$^2$ COSMOS \citep{COSMOS_overview2007} and XMM-LSS \citep{XMM_2_2008ApJS..176....1F,XMM_3_2008ApJS..179..124U} fields, along with forced photometry from the Subaru Hyper-Suprime Cam observations in these fields \citep[e.g.,][]{Subaru_HSC_SSP_Overview,Suabru_HSC_SSP_1,aihara_second_2019}. We then selected LAE candidates based on their narrow-band filter flux, flux excess relative to the broad-band or other narrow-band filters, and (where there exist broad-band detections) a color cut to exclude red low-redshift galaxy interlopers. Transmission curves for relevant filters can be seen in Figure~\ref{fig:filters}.

DESI is a medium-resolution ($R\equiv\lambda/\Delta\lambda=2000-5300$) $\approx$5000-fiber multi-object spectrograph \citep{DESI_Instrument_Overview, Corrector.Miller.2023, SurveyOps.Schlafly.2023,DESI_Target_Selection_Pipeline, FiberSystem.Poppett.2024} dedicated to carrying out a 5-year survey of over 40 million extragalactic sources to measure the expansion history of the universe and constrain the dark energy equation of state \citep[for details see][]{DESI_Science_Overview2016, DESI_Survey_Validation, DESI2024.VII.KP7B, desicollaboration2025datarelease1dark}. The data were reduced using the DESI spectroscopic pipeline \citep{DESI_Spectro_Pipeline}.

The final catalog consists of 11,634 LAE candidate spectra that have been visually inspected (VI), each of which includes a redshift estimate, an assessment of redshift quality between 0 and 4 (where 0 indicates no detected flux and 4 indicates a firm redshift estimate confirmed by multiple lines), and any comments about the object. We remove any spectra with VI quality $<$ 2.5, and spectra with comments that contain the phrase ``QSO" or ``AGN" to ensure a high-quality sample of LAEs. We additionally remove duplicates and any galaxies with \delLAE{} = -1. These cuts yield 2,831 spectra of galaxies in the N419 redshift range ($2.4 < z < 2.5$) and 1,937 galaxies in the N501 redshift range ($3.07 < z < 3.18$). 

Each galaxy in this sample has a redshift determined by visual inspection- however, while these are often accurate, they can be imprecise due to variations in the identification of the center of the \lya{} line between different volunteers. This is especially true in the case of double-peaked \lya{} line profiles, where identifying the redshift based on either peak or their center could be a reasonable guess.

To remove this potential source of uncertainty, we use the automated  pipeline from \cite{uzsoy_bayesian_2025} to determine redshifts for these spectra. This method uses Marginalized Analytic Dataspace Gaussian Inference for Component Separation \citep[MADGICS;][]{Saydjari+23, MADGICS}, a Bayesian component separation method, to decompose LAE spectra into sky residual, LAE, and residual components. The spectroscopic redshift is determined by optimizing the change in $\chi^2$ from adding an LAE component at a given redshift. As with VI redshifts, this pipeline assigns redshifts based on the peak of the identified \lya{} line, which is often offset from the systemic redshift. 

\begin{deluxetable}{cccccc}[h]
\tabletypesize{\footnotesize}
\tablewidth{0pt}

 \tablecaption{Number of galaxies with spectra in vs. out of protocluster (PC) in each NB \& field. \label{tab:num_galaxies}}

 \tablehead{
 \colhead{NB} & \colhead{Redshift} & \colhead{Field} & \colhead{\# in PC} & \colhead{\# not in PC} & \colhead{Total} }

\startdata
N419 & $2.4 - 2.5$ & XMM-LSS & 91 & 934 & 1025\\
N419 & $2.4 - 2.5$ & COSMOS & 79 & 1737  & 1816\\
N501 & $3.07 - 3.18$ & COSMOS & 146 & 1791 & 1937\\
\enddata
\end{deluxetable}

Previous work has shown that redshifts determined with this pipeline agree with VI redshifts to within 0.005 for $>$90\% of LAE spectra tested \citep{uzsoy_bayesian_2025}. In the case of double-peaked lines, the pipeline will assign redshift based on the larger peak. The use of an automated procedure for redshift determination ensures uniformity in treatment of double-peaked sources, which is important for line profile studies.

DESI spectra cover the observed wavelength region between 3600 and 9800 \AA\ with 0.8-\AA\ bins. Once the redshift is determined for each galaxy, the spectra are shifted to rest-frame and interpolated onto a wavelength range of 500-3000 \AA\ with 0.2-\AA\ bins using Lanczos interpolation. The spectra are not normalized.

\begin{figure*}[ht!]
\epsscale{1.15}
\plottwo{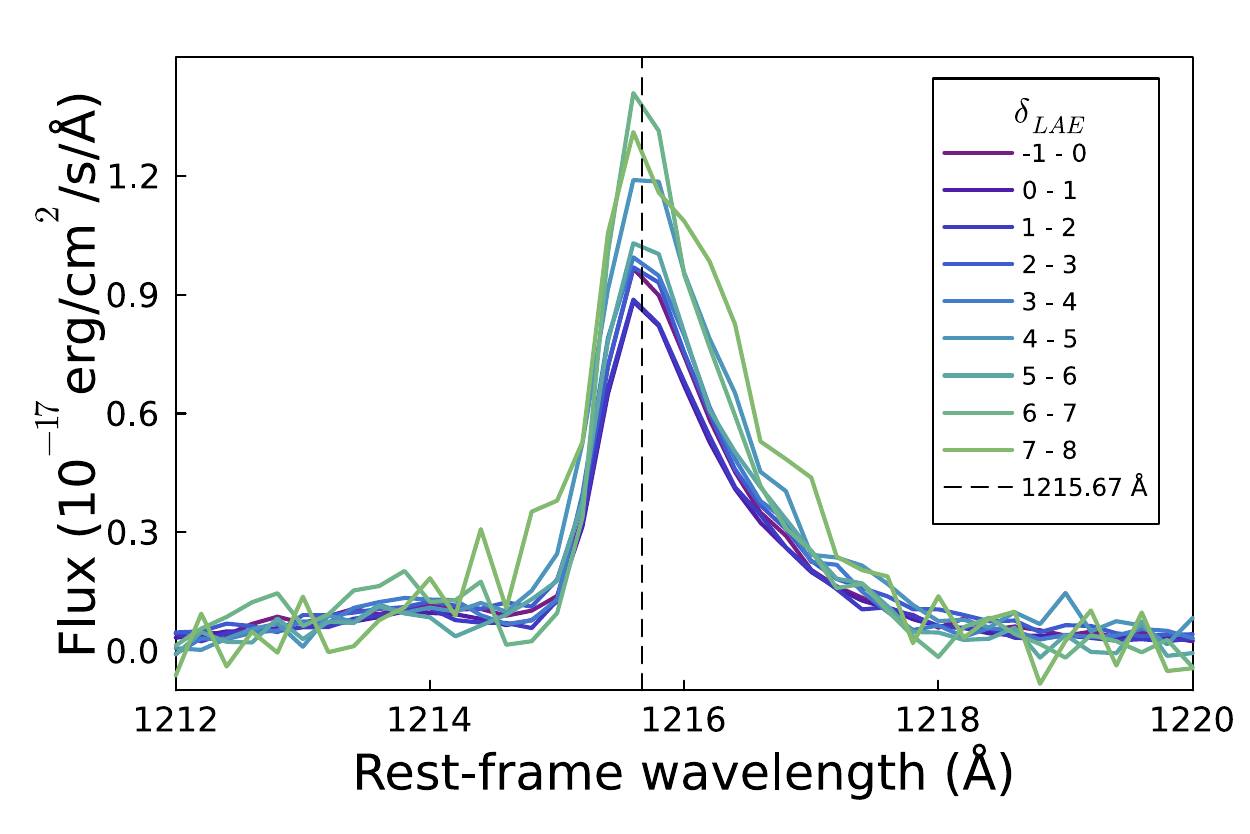}{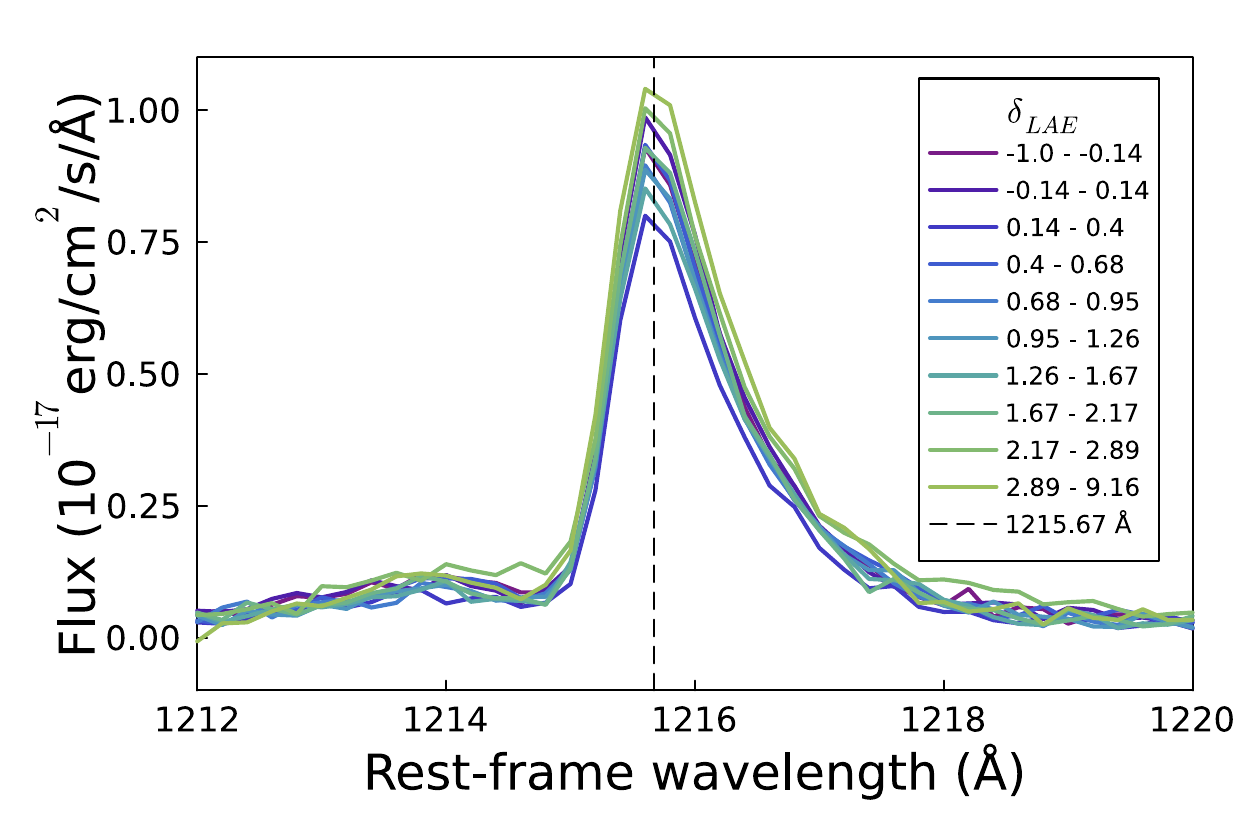}
\caption{Inverse-variance-weighted mean stacks of the \lya{} line profile for DESI/ODIN LAEs in the N501 band, binned in linear bins (left) and percentile bins (right) of \delLAE. Lighter colors indicate denser environments. The legend denotes the range of \delLAE{} values represented in each bin. The dashed line represents 1215.67 \AA, the rest-frame \lya{} wavelength. Bins containing only one galaxy are excluded.}
\label{fig:501_stacks}
\end{figure*}

\subsection{LAE Density Estimates}

The ODIN LAE surface density maps are created using the Voronoi tessellation of the distribution of LAEs in a given field \citep{ramakrishnan+24}. To apply this method, the field is divided into cells containing the location of one LAE, where the cell encloses the area of points closest to that LAE than any other. The surface density $\Sigma_{\rm LAE}$ at every point is then the inverse of the area of its associated Voronoi cell \citep{ramakrishnan+24}. 

The LAE surface overdensity parameter \delLAE{} is defined as:

\begin{equation}
    \delta_{\rm LAE} = \frac{\Sigma_{\rm LAE}}{\langle \Sigma_{\rm LAE} \rangle} - 1
\end{equation}
where $\Sigma_{\rm LAE}$ denotes the local LAE surface density and $\langle \Sigma_{\rm LAE} \rangle$ denotes the mean LAE surface density \citep{ramakrishnan+24}.

In addition to deriving the continuous \delLAE{} values, \cite{ramakrishnan_odin_2025} also identify specific protocluster regions throughout the field of view. The criteria for identifying protocluster regions include spanning a minimum area of 40 cMpc$^2$ and having \delLAE{} $>$ 2.6 for galaxies observed with the N501 band and $>$ 2.3 for the N419 band. The lower-redshift band has a lower threshold due to the higher number density of LAEs at z $\approx$ 2.45 \citep{ramakrishnan_odin_2025}.

The surface density maps from \cite{ramakrishnan_odin_2025} are then convolved with a fine, uniformly sampled grid, and each galaxy is assigned the \delLAE{} value of its nearest point in (RA, DEC) in the density grid. Note that this is only two-dimensional; in reality, protoclusters may not span the entire redshift range. The values of \delLAE{} range from -1 to 6.17, 8.89, and 9.16 for the N419 COSMOS, N419 XMM-LSS, and N501 COSMOS datasets, respectively. A galaxy is determined to be in a protocluster if its nearest point in the density grid is in a protocluster region. Table~\ref{tab:num_galaxies} shows the number of galaxies in this dataset that are in vs. out of protoclusters in each narrow band and field of view. We exclude any galaxies with \delLAE{} = -1, which denotes that it is out of the bounds of the broadband data.


\section{Results} \label{sec:results}

\subsection{First results}
\label{sec:firstresults}

To evaluate the effect of local environmental density on the shape of the \lya{} line profile, we stack spectra with similar \delLAE{} values and compare stacks between different local densities. Figure~\ref{fig:501_stacks} shows inverse-variance weighted mean stack of the \lya{} line profile for LAEs grouped in bins of \delLAE{}. The left panel of Figure~\ref{fig:501_stacks} shows the stacks in linear bins of \delLAE{}, where the stacks from bins that contain a single galaxy are not shown. The data suggest that galaxies in denser environments tend to have larger \lya{} line luminosity and EW compared to galaxies in less dense environments.

\begin{figure}[ht!]
\epsscale{1.15}
\plotone{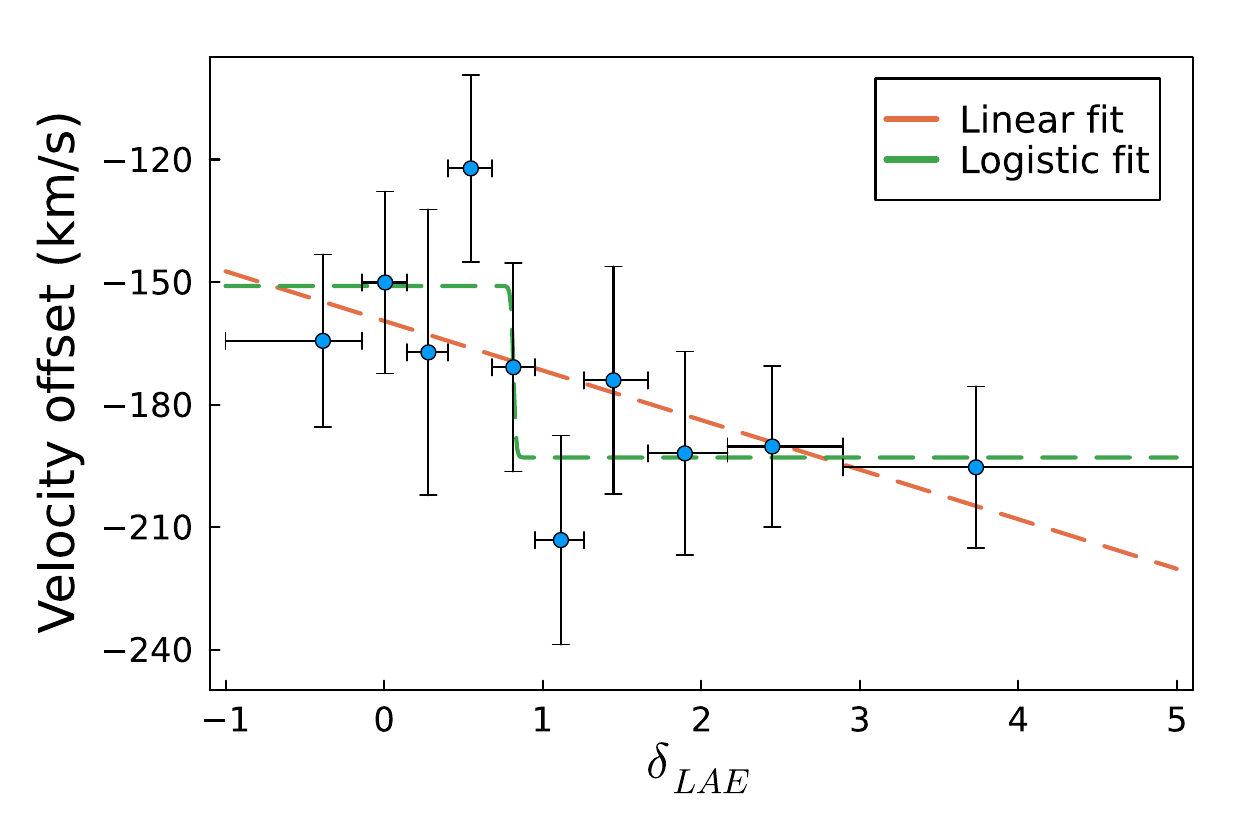}
\caption{Velocity offset of the \lya{} redshift from the systemic redshift (determined from He II emission), as a function of LAE overdensity \delLAE{}. Offsets were determined from the tenth percentile-binned stacks seen in Figure~\ref{fig:501_stacks}. Horizontal error bars denote the range of \delLAE{} values in each stack, while vertical error bars denote 1$\sigma$. Dashed lines denote linear (orange) and logistic (green) fits.}
\label{fig:velocity_offsets}
\end{figure}

\begin{figure*}[ht!]
\epsscale{1.15}
\plottwo{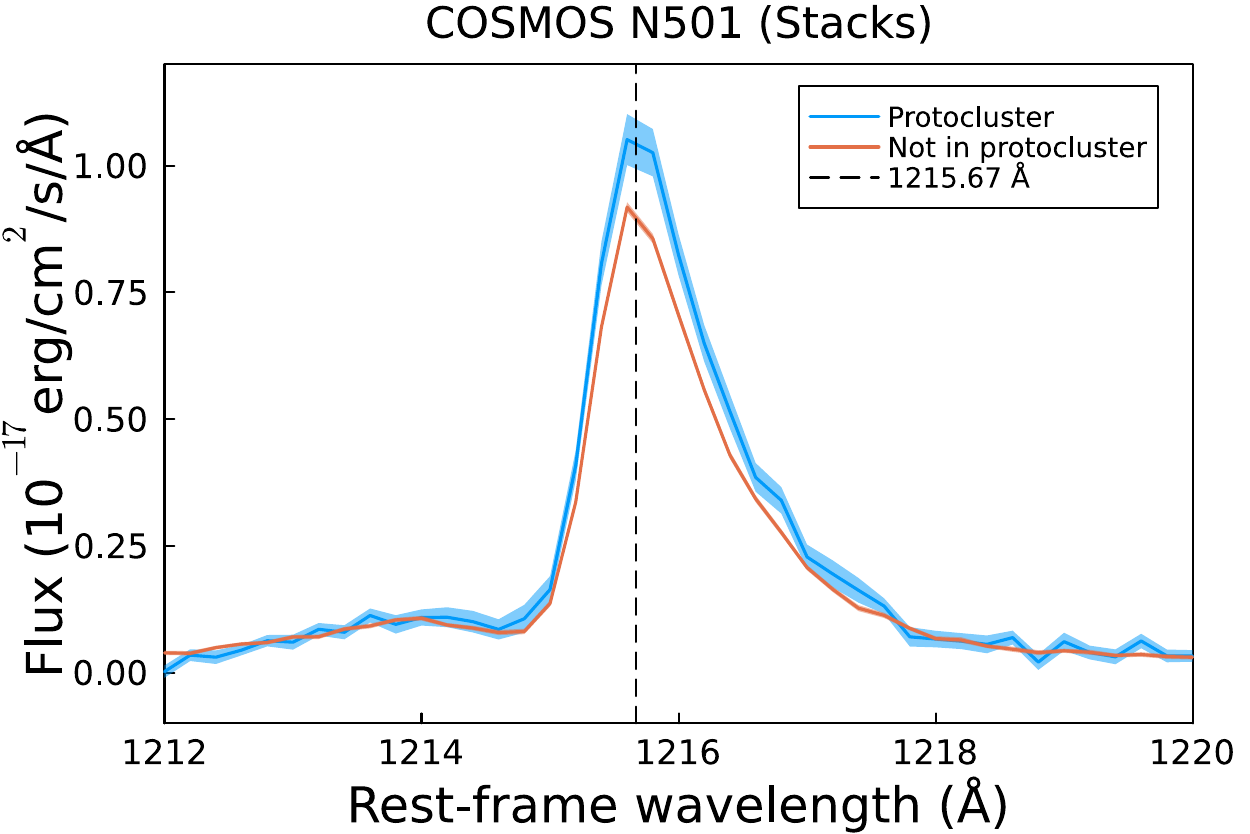}{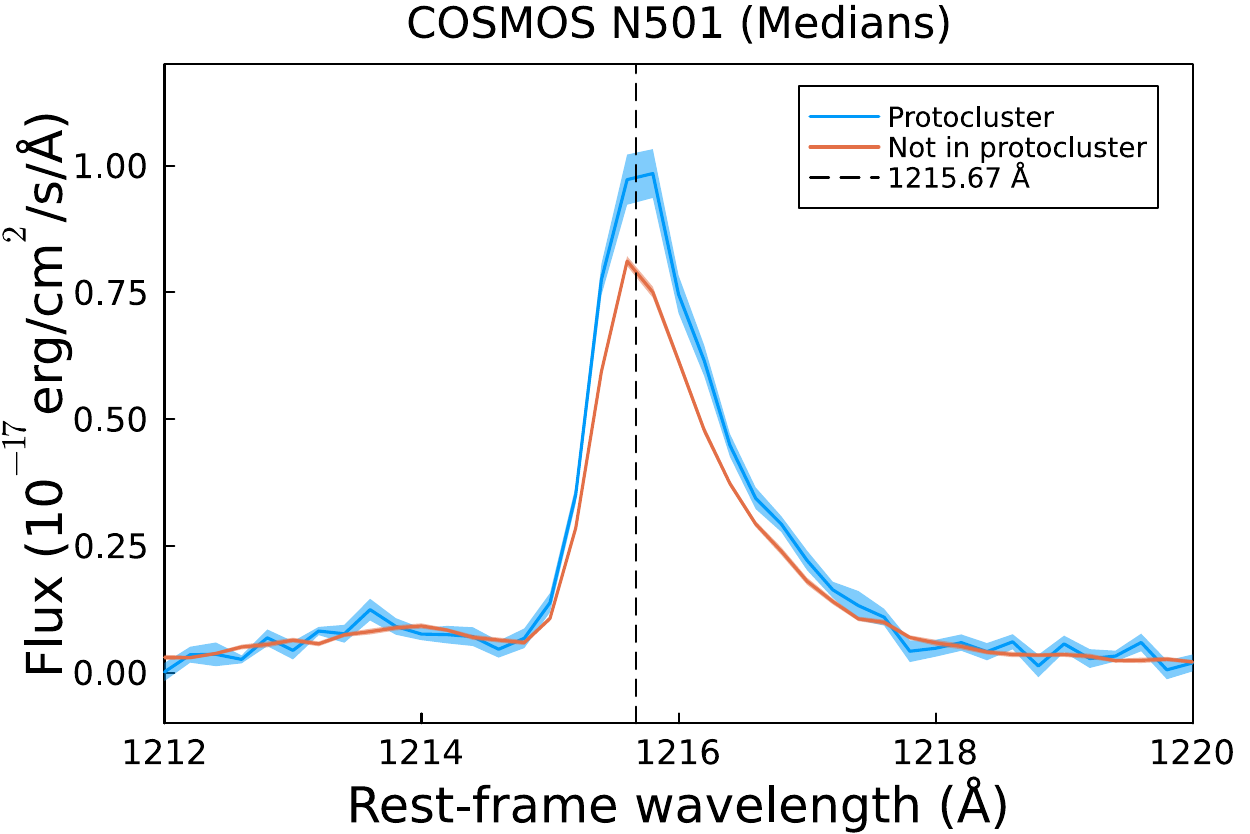}
\caption{Inverse-variance-weighted mean stacks (left) and median spectra (right) of the Ly$\alpha{}$ line profile for DESI/ODIN LAEs in the N501 band, inside (blue) and outside (orange) of protocluster regions in the COSMOS field. The dashed line represents 1215.67 \AA, the rest-frame \lya{} wavelength. Shaded regions denote 1$\sigma$ bootstrapped errors.}
\label{fig:501_proto_stacks}
\end{figure*}

Protocluster regions are uncommon and small compared to the size of the entire field, and thus there are many more galaxies in our sample in regions with lower values of \delLAE{} than with higher values (see Table~\ref{tab:num_galaxies}). In the left panel of Figure~\ref{fig:501_stacks}, there are 718 galaxies in the most populous bin with \delLAE{} between 0 and 1, while there are only four galaxies in the highest bin with \delLAE{} values between 7 and 8. As a result, the stacks become noticeably noisier as the values of \delLAE{} increase. 

To mitigate the effects of sample noise, we rebin the data in tenth-percentile bins of \delLAE{}, with $\sim$193 galaxies in each stack, seen in the right panel of Figure~\ref{fig:501_stacks}. The same trend can be seen as in the left panel, with galaxies in denser environments having higher \lya{} line luminosity and larger EWs than galaxies in the less dense environments. These less noisy stacks also allow us to better probe the rest of the shape of the line profiles- the stacks are almost exactly the same on the blue side, with the width of the line increasing mostly on the red side.

We additionally investigate the velocity offset of the \lya{} redshift from the systemic redshift as a function of \delLAE{}. To probe this, we jointly fit three Gaussians to the He II and OIII] doublet, where the wavelength ratios between the lines and the width of the Gaussians is fixed and only the He II wavelength is optimized. We perform this analysis on the tenth percentile-binned stacks (shown in the right panel of Figure~\ref{fig:501_stacks}) in order to reduce the effects of noise. 

The results, seen in Figure~\ref{fig:velocity_offsets}, show a velocity offset ranging between -120 and -240 km/s, with a slight downward trend with increasing \delLAE{}. A linear fit to the points yields $\Delta v = -12.14 \delta_{\rm LAE} -159.51$ and a logistic fit yields asymptotic values of approximately -151 km/s and -193 km/s at low and high environmental densities, respectively. The observed trend can potentially be attributed to more absorbing gas in galaxies in dense regions, which could offset the \lya{} peak center. Our values are in relative agreement with \cite{herreraalcantar2025}, who determined an LAE velocity offset of -241 $\pm$ 20 km/s relative to \lya{} forest absorption lines in the same vicinity in DESI spectra. 

Figure~\ref{fig:501_proto_stacks} shows both the \lya{} line profiles for the LAEs observed in the N501 band in the COSMOS field, inside (blue) and outside (orange) of protocluster regions. To validate our result, we show the \lya{} line profiles as both inverse-variance-weighted mean stacks and median spectra. Shaded regions show 1$\sigma$ bootstrapped errors derived from 50 repetitions for all figures in this work. Here we see the same trends as in Figure~\ref{fig:501_stacks}, where galaxies in protoclusters have larger \lya{} line luminosity than galaxies outside of protoclusters.

To quantify the difference in line luminosities between the in- and out-of protocluster stacks, we compare the integrated line luminosities. To calculate this, for each stack, we estimate the continuum flux as the average flux value between 1250-1450 \AA. We then integrate the spectrum between 1209.8-1219.8 \AA, subtracting out the continuum flux from the red half of the \lya{} line (spanning 1215.67-1219.8 \AA). Our resulting estimates agree well with those derived photometrically by \cite{nagaraj_odin_2025}. We then compare the integrated line luminosities of the in- vs. out-of protocluster samples: their ratio is 1.14 for the stacked spectra and 1.18 for the median spectra. 

\begin{figure*}[ht!]
\epsscale{1.15}
\plottwo{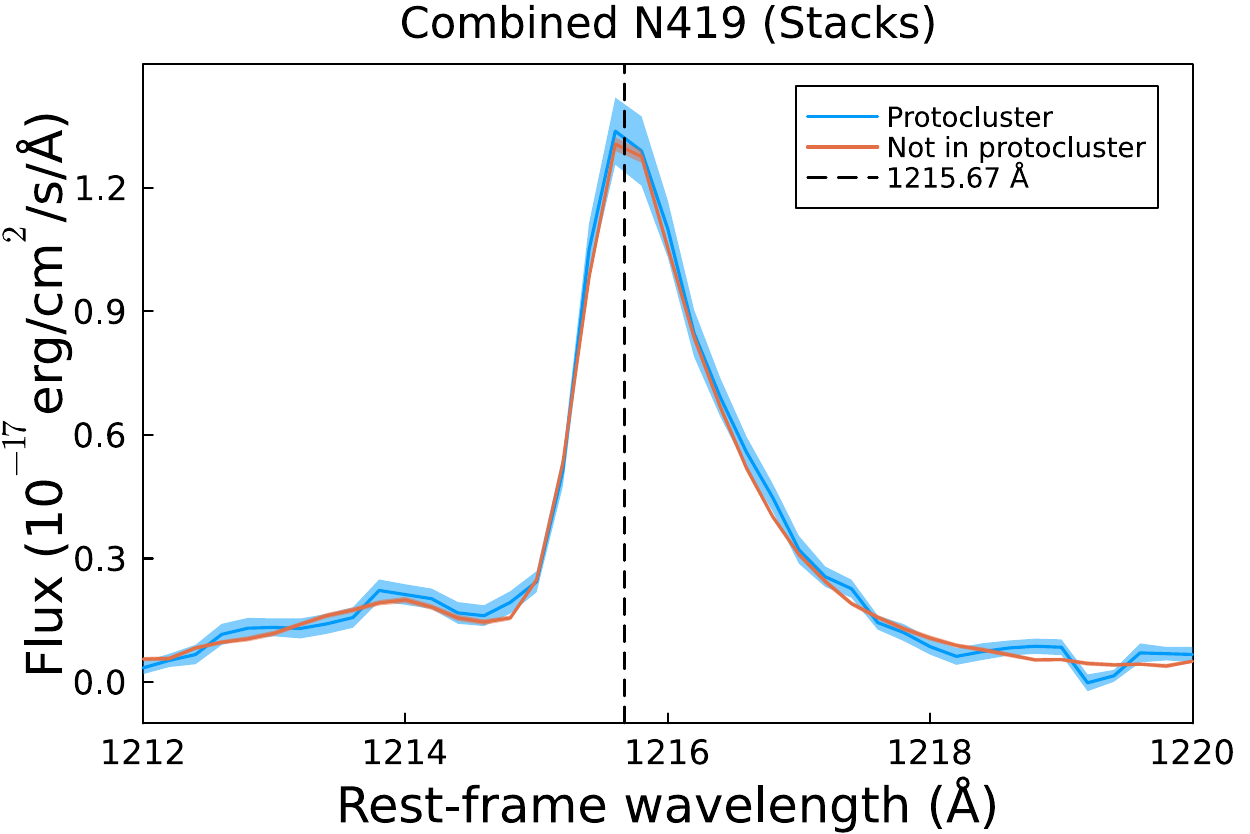}{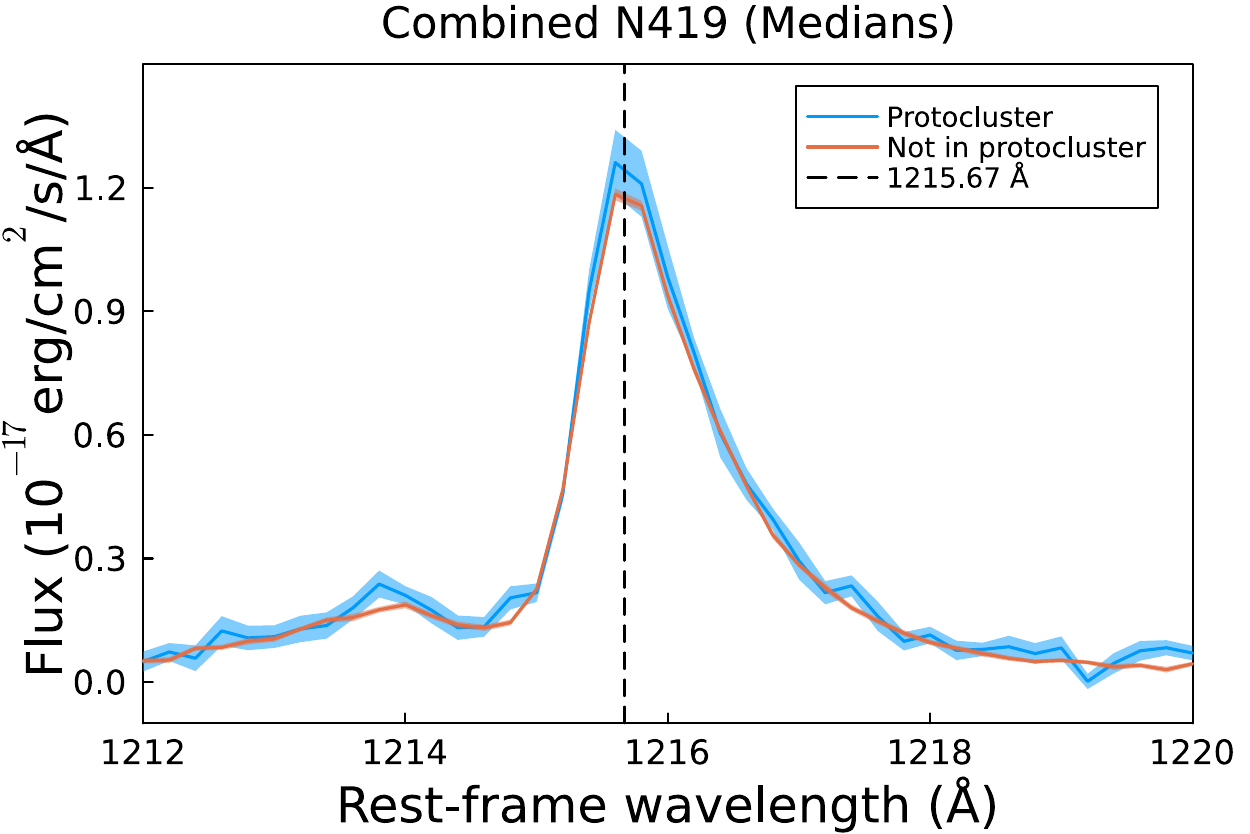}
\caption{Inverse-variance-weighted mean stacks (left) and median spectra (right) of the Ly$\alpha{}$ line profile for DESI/ODIN LAEs in the N419 band, inside (blue) and outside (orange) of protocluster regions combined over the COSMOS and XMM-LSS fields. The dashed line represents 1215.67 \AA, the rest-frame \lya{} wavelength. Shaded regions denote 1$\sigma$ bootstrapped errors.}
\label{fig:combined_419_proto_stacks}
\end{figure*}

Figure~\ref{fig:combined_419_proto_stacks} shows the same stacks and median spectra inside and outside of protocluster regions, but for galaxies observed with the N419 (z $\approx$ 2.45) band combined across the COSMOS and XMM-LSS fields. In the N419 band we do not see the same strong trend as in the N501 data; the inside and outside protocluster lines are within the error regions. In each field, there are significantly fewer spectra of galaxies in protocluster regions than in the N501 COSMOS dataset, so we combine them to increase our statistical power (separate stacks from each field can be seen in Appendix~\ref{appendix:separate_Stacks}). Nonetheless, comparison between the line profiles inside and outside of protoclusters suggests that there may be increased star formation in overdense regions of LAEs at z $\approx$ 3.12, but not at z $\approx$ 2.45. 


\subsection{Double-peaked sources}
\label{sec:double_peaks}

Double-peaked \lya{} profiles are relatively common, and can be caused by resonant scattering of \lya{} photons, outflows, interacting systems, geometry of ionized gas around the galaxy, IGM transmission, or many other physical processes \citep{ouchi_observations_2020, hayes_spectral_2021,almadamonter_crossing_2024, vitte_muse_2025}. We identify double-peaked sources in the COSMOS N501 dataset by selecting galaxies based on their coefficients of specific principal components of the \lya{} line (see Appendix~\ref{appendix:double_peaks} for more details). Out of the 1,937 galaxies in this dataset, we identify 312 (16.1\%) as having double-peaked line profiles, with 28 (9\%) of them in protocluster regions. As a point of comparison, 320 (31.2\%) galaxies in the N419 XMM-LSS dataset and 659 (36.3\%) galaxies in the N419 COSMOS dataset are identified as double-peaked, suggesting redshift evolution in double-peak fraction, possibly due to IGM transmission \citep{hayes_spectral_2021}. The literature suggests LAE double-peaked fractions are likely between 30 and 50\%, but also that they could decrease with redshift \citep{Kulas+12, Trainor+15, Kerutt+22, vitte_muse_2025, mukherjee2025}. 7.5\% of all of the observed galaxies (including single- and double-peaks) in the COSMOS field at redshift 3.12 are in protocluster regions, suggesting that double-peaked galaxies are not preferentially present in denser or sparser regions.

\begin{figure}[ht!]
\epsscale{1.2}
\plotone{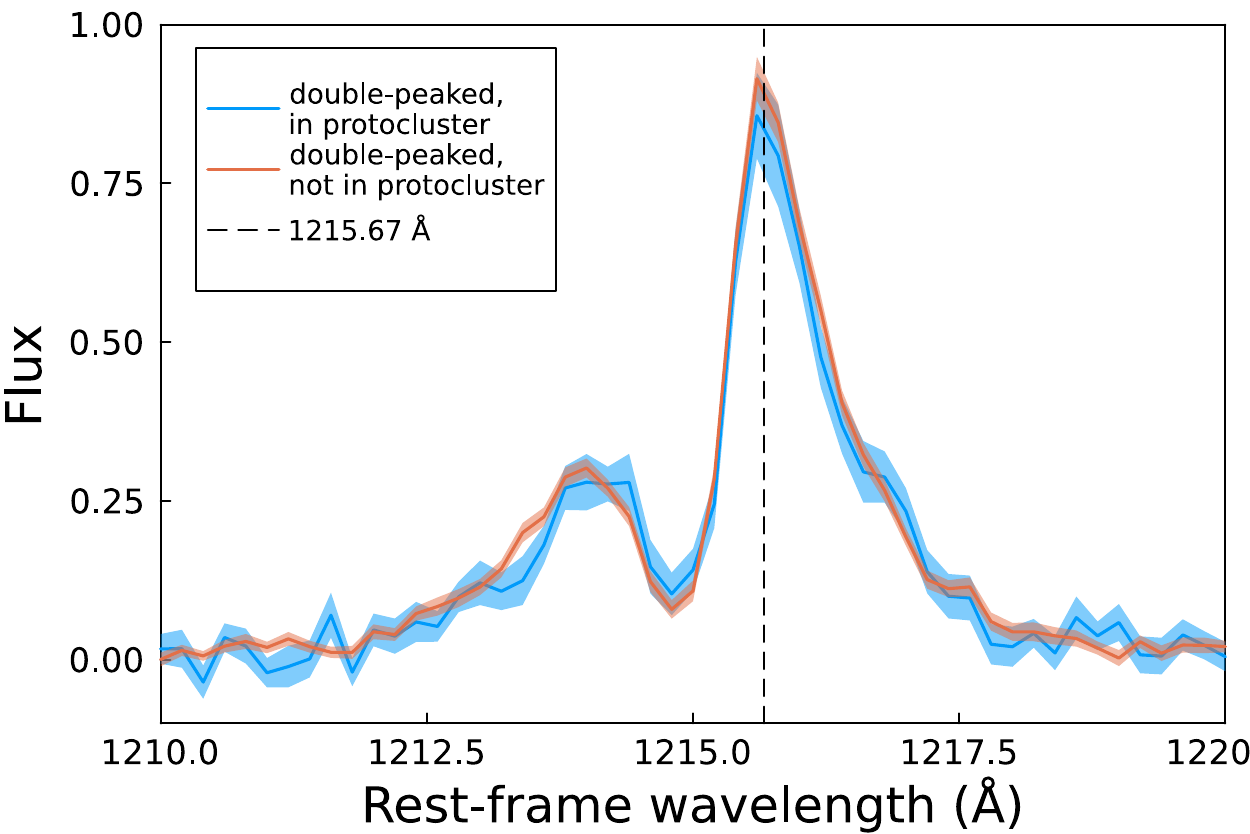}
\caption{Inverse-variance-weighted mean stacks of the Ly$\alpha{}$ line profile for double-peaked DESI/ODIN LAEs inside vs. outside of protocluster regions in the COSMOS N501 dataset. Shaded regions denote 1$\sigma$ bootstrapped errors.}
\label{fig:double_stacks}
\end{figure}

Figure~\ref{fig:double_stacks} shows the same analysis as in Figure~\ref{fig:501_proto_stacks}, stacks inside and outside of protocluster regions, but only for these double-peaked sources. There is no discernable difference between the \lya{} luminosity between in- and out-of-protocluster double-peaked galaxies, suggesting that there is overlap between environmental density and the myriad physical processes that cause a double-peaked line profile. These could include the geometry of the gas in the galaxy, which is likely different in overdense regions, and resonant scattering of \lya{} photons, which is likely more common in regions with more ionized and neutral gas.


\subsection{Investigation of potential selection effects}

\begin{figure}[ht!]
\epsscale{1.2}
\plotone{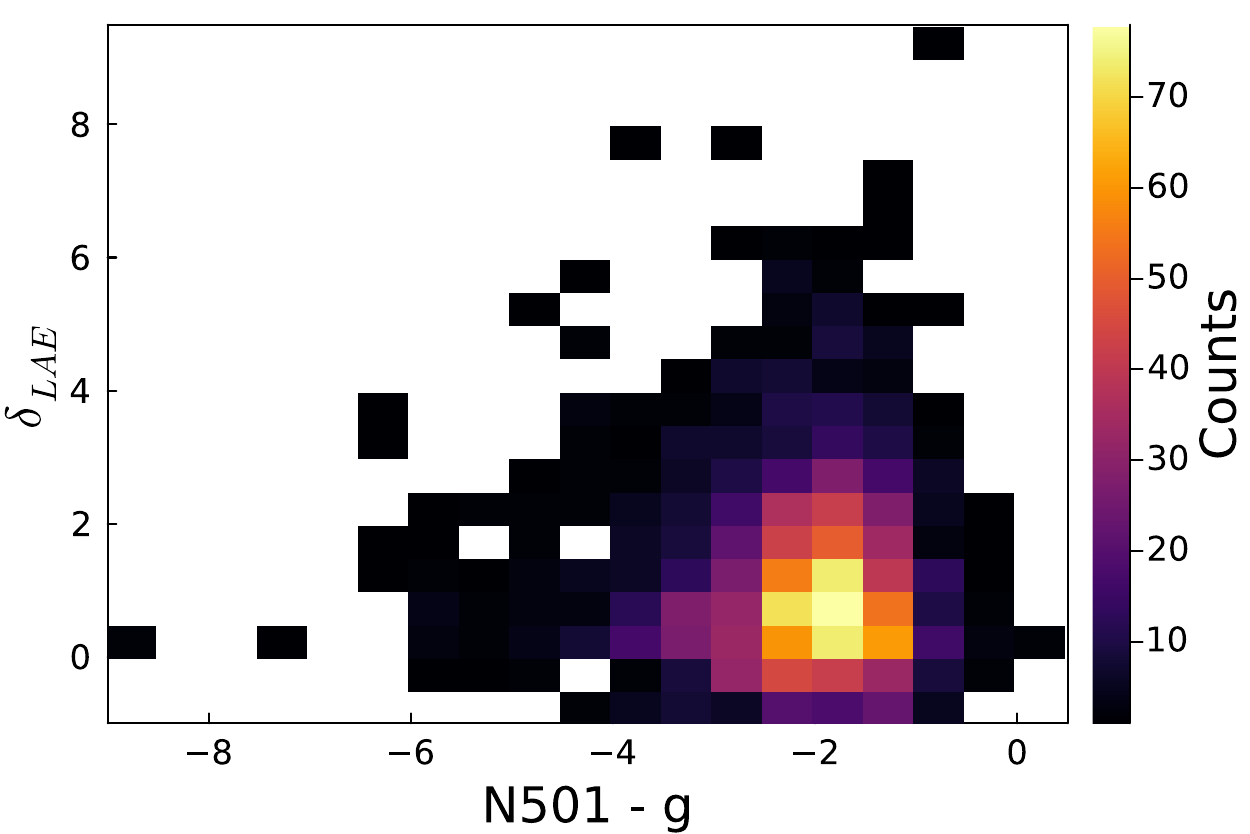}
\caption{Local density \delLAE{} vs. NB excess magnitude (N501 - $g$) for 1,937 galaxies observed with the N501 filter in the COSMOS field.}
\label{fig:mag_delta}
\end{figure}

One potential explanation of the observed difference in \lya{} line luminosity inside vs.\ outside of protocluster regions is that it is a selection effect, where brighter galaxies in overdense regions are more likely to be observed. LAEs in ODIN are selected by their excess magnitude in the NB compared to the broadband magnitude (using bands shown in Figure~\ref{fig:filters}). Figure~\ref{fig:mag_delta} shows the distribution of NB excess magnitude over \delLAE{}. There is little evidence of an intrinsic relationship between NB excess and environmental density; \cite{dey_spectroscopic_2016} conduct a similar analysis in their Figure 11 and find a slight trend amid large scatter, but this is likely too weak to cause the significant luminosity difference between in- and out-of-protocluster galaxies seen in Figure~\ref{fig:501_proto_stacks}. The lack of a relationship between environmental density and \lya{} luminosity has also been suggested using the TNG simulations \citep{andrews+25}.

\begin{figure*}[ht!]
\epsscale{1.2}
\plotone{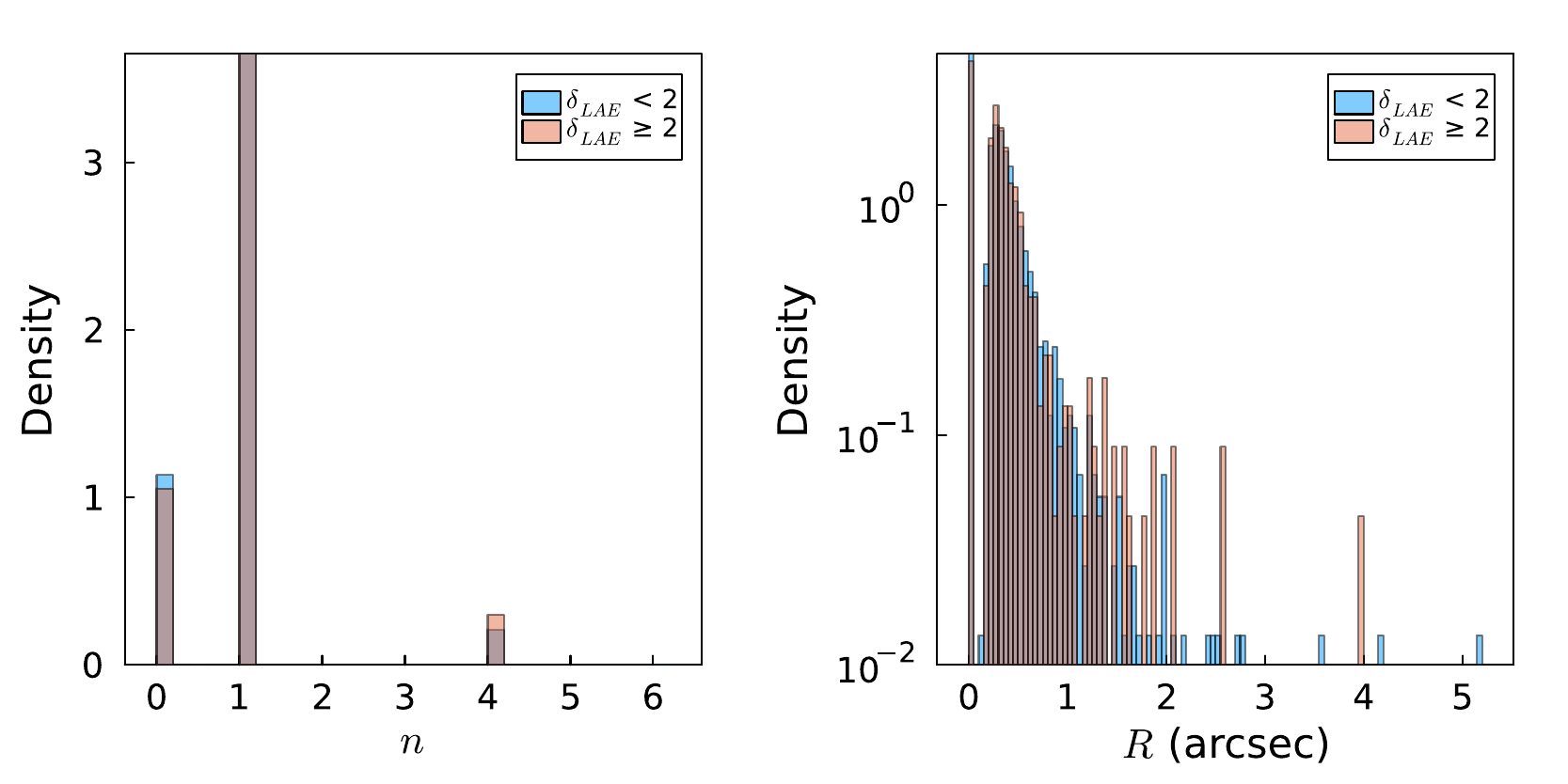}
\caption{Distributions of the Sersic parameter $n$ (left) and half-light radius $R$ (right) for COSMOS N501 LAEs for \delLAE{} $\geq$ 2 (orange) and \delLAE{} $<$ 2 (blue). }
\label{fig:sersic}
\end{figure*}

Variations in galaxy morphology as a function of environment can also result in the line profiles measured in a DESI fiber. For example, it is possible that a DESI fiber does not span the entire galaxy but instead only points at its center, and if one set of galaxies has a luminosity distribution that is more concentrated at the center, then this could lead to artificially higher observed line luminosities in their spectra. Figure~\ref{fig:sersic} shows the distribution of the Sersic parameter $n$ and half-light radius $R$ of a Sersic profile \citep{sersic} for all N501 COSMOS galaxies, with populations split on threshold value of \delLAE{} = 2. The distributions of both parameters are quite similar between the two populations and agree with values from the literature \citep{shibuya_morphologies_2019}, suggesting that there is no significant difference in the morphology of galaxies in over- or under-dense environments at these redshifts. 

The enhanced line luminosity inside of the protocluster regions could also potentially be because these galaxies have higher continuum flux values. Bias towards higher redshifts could also produce this effect, as emission lines that are partially outside of the NB wavelength range would have to be stronger to still be selected based on NB excess flux. We can investigate this effect by selecting out-of-protocluster galaxies that most closely resemble the population of in-protocluster galaxies in $r$ band magnitude and redshift, and verifying that the effect is unchanged. 

\begin{figure*}[ht!]
\epsscale{1.15}
\plottwo{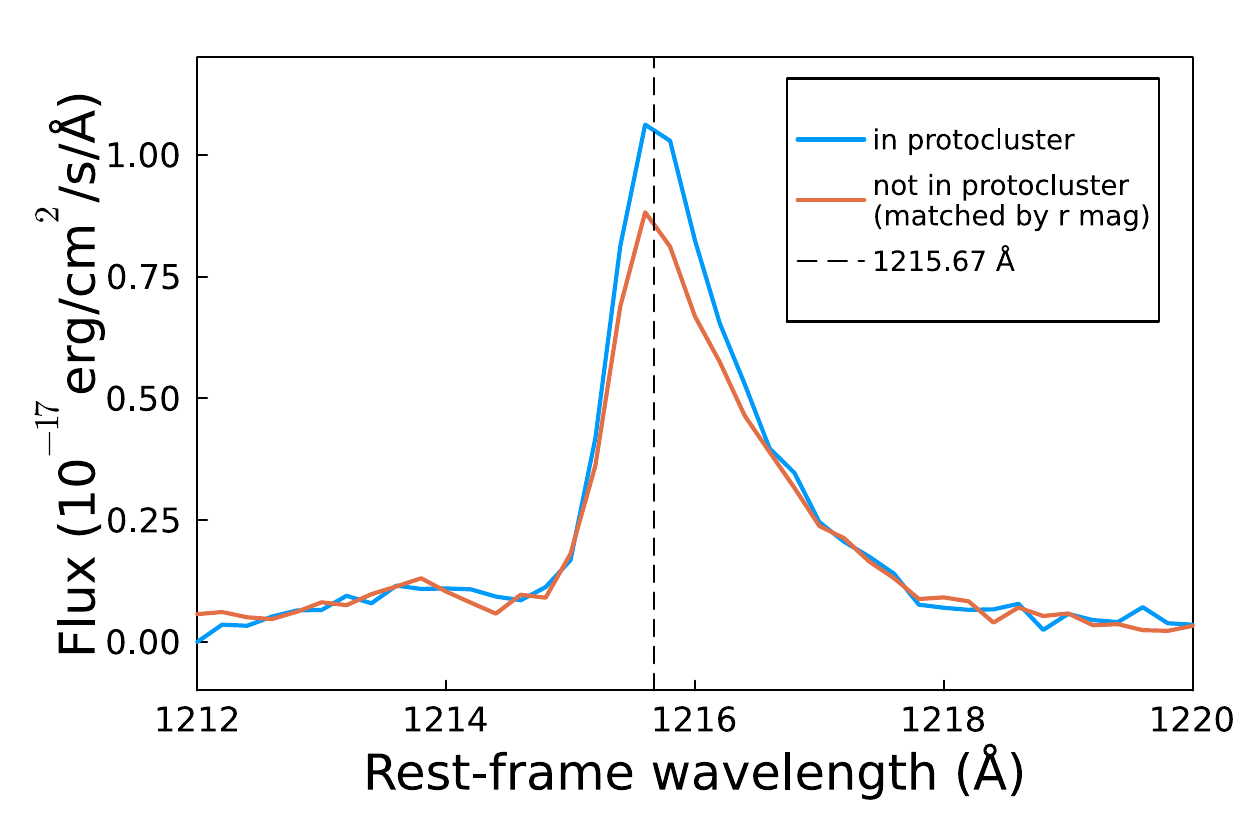}{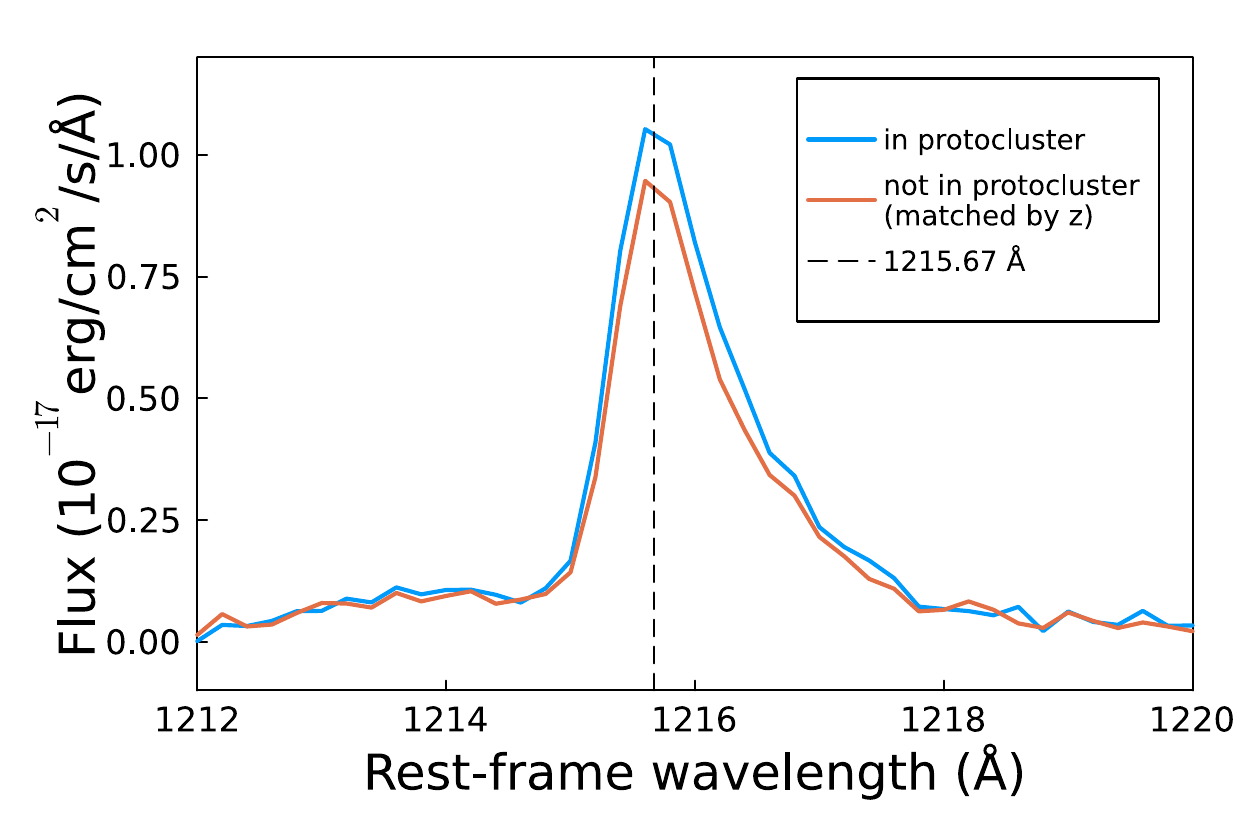}
\caption{Inverse variance-weighted-mean stacks of COSMOS N501 LAEs inside of protocluster regions (blue) and outside of protocluster regions(orange), matched to the in-protocluster galaxies by $r$ band magnitude (left) and redshift (right).}
\label{fig:match}
\end{figure*}

For each of the 146 galaxies selected with the N501 band that are in protocluster regions in the COSMOS field, we select the out-of-protocluster galaxy with the closest value of $r$ band magnitude (see Figure~\ref{fig:filters} for transmission curves) and redshift (these need not be the same galaxy). Figure~\ref{fig:match} shows the inverse-variance weighted mean stacks of the \lya{} line inside and outside of the protocluster (as in the left panel of Figure~\ref{fig:501_proto_stacks}), but now comparing with these new selected ``matched" populations (the blue curves are exactly the same as in the left panel of Figure~\ref{fig:501_proto_stacks}). The out-of-protocluster stacks remain largely unchanged, and the significant difference in line luminosity remains visible. This analysis allows us to essentially control for redshift and continuum flux, and shows that the effect remains even when these parameters are accounted for. This suggests that the elevated \lya{} line luminosity can likely be attributed to the local density of the galaxies instead of any other variables.

There are potential downstream effects of the increased \lya{} line luminosity inside protocluster regions. LAEs are targeted with their narrow-band excess flux, so LAEs in protocluster regions are more likely to be observed given their elevated flux. For the COSMOS N501 field, 146 / 1937 = 7.5\% of LAEs are in protocluster regions. However, if we select a flux-limited sample of galaxies with integrated line luminosities above 1 $\times 10^{-16}$ erg/cm$^2$/s, 40 of the resulting 376 galaxies (10.6\%) are in protocluster regions. However, this effect is relatively small, and the majority of galaxies observed with ODIN are not in protocluster regions, so this dataset still allows for robust characterization of in vs. out of protocluster LAEs.

\section{Discussion} \label{sec:discussion}

We see evidence of a relationship between the \lya{} line profile and environmental density in the N501 dataset ($z \approx 3.12$), and potential evidence of a subtler effect in N419 ($z \approx 2.45$) LAEs from ODIN. Figure~\ref{fig:501_stacks} shows that the \lya{} line luminosity and EW climb when approaching denser environments, while Figure~\ref{fig:501_proto_stacks} shows a significant difference when comparing LAEs inside vs. outside of protocluster regions. The increased \lya{} emission in these overdense regions in the N501 band suggests that there is more star formation than in lower-density regions.

Previous work from the ODIN collaboration shows that the \lya{} luminosity function is elevated inside of protocluster regions in all three narrow bands, but notes that the amount of this elevation increases with redshift \citep{nagaraj_odin_2025}. The reduced effect in the N419 galaxies could be in part from the known reversal of the SFR-density relationship, with galaxies at higher redshifts having more star formation in more dense environments \citep{elbaz_reversal_2007, Lemeaux+22}. Ongoing surveys that sample larger volumes over wider redshift ranges will allow for most robust investigation of this effect.

\begin{figure}[ht!]
\epsscale{1.2}
\plotone{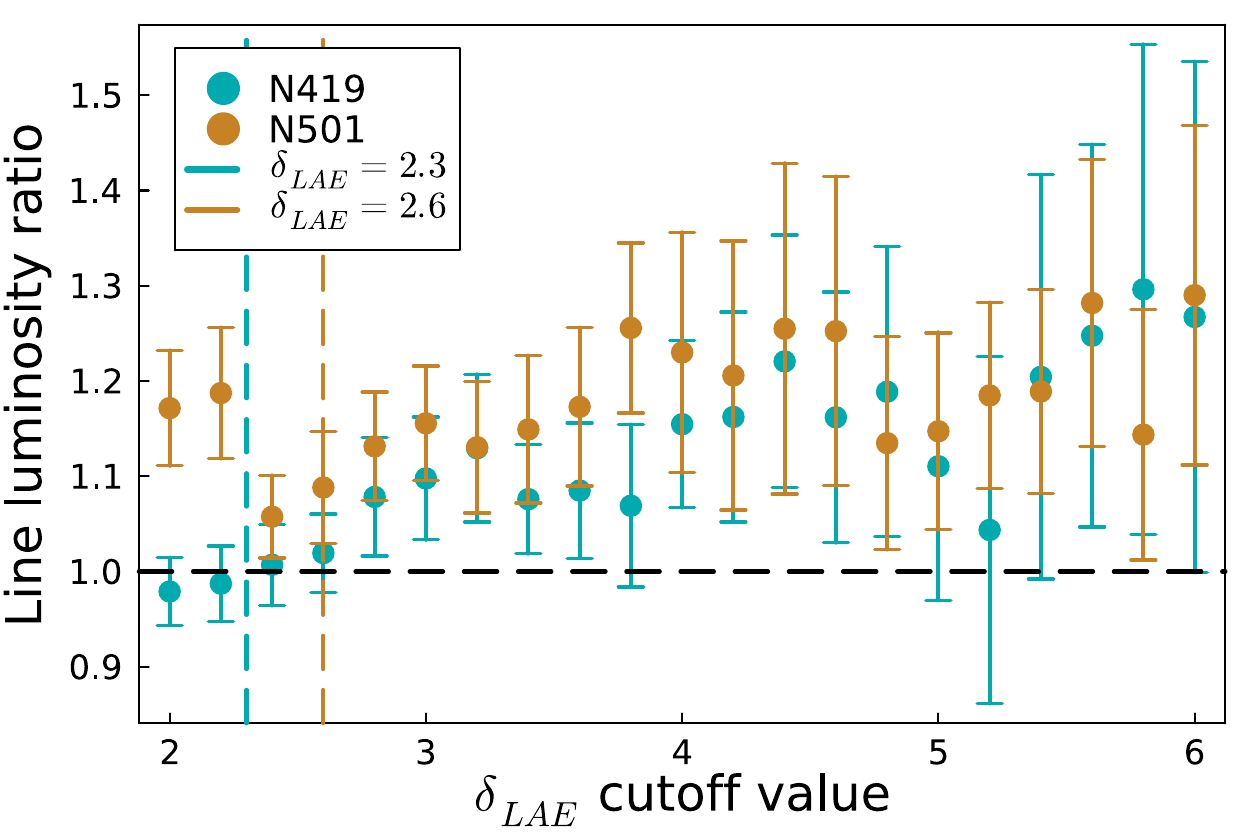}
\caption{Ratio of integrated \lya{} line flux inside and outside of protocluster regions defined by a varying cutoff \delLAE{} value for N419 galaxies (teal) and N501 galaxies (brown). Error bars show bootstrapped errors. Dashed lines denote \delLAE{} = 2.3 and 2.6, the cutoff values used by \citep{ramakrishnan+24}, in addition to a minimum area threshold, to denote protocluster regions in the N419 (z $\sim$ 2.45) and N501 (z $\sim$ 3.12) NBs, respectively.}
\label{fig:line_ratios}
\end{figure}

This work approximates a narrow-band field as two-dimensional, and galaxies are assigned as in- or out- of protocluster regardless of their position along the line of sight. In reality, protocluster regions likely do not span the entire redshift range probed by the NB filter. By comparing 3D simulations and their 2D projections, previous work has validated that this approximation is adequate for the ODIN survey \citep{andrews+25}. However, this work could be improved by including 3D information, and future work spanning larger redshift ranges will need to consider this additional dimension.

The comparison of line profiles in Figures~\ref{fig:501_proto_stacks} and \ref{fig:combined_419_proto_stacks} depends on how a protocluster region is defined. While the choice of protocluster overdensity cutoff values of \delLAE{} = 2.3 for N419 and 2.6 for N501 galaxies is well-motivated by \cite{ramakrishnan_odin_2023}, the strength of the observed difference in line profiles in vs. out of protocluster varies with this cutoff value.  To probe this relationship, we calculate the integrated line flux ratio, hereafter referred to as $\eta$, based on \delLAE{} cutoff values ranging from 2 to 6. For each cutoff value, the spectra of the populations above and below the cutoff are stacked, and the rest-frame line profile is integrated between 1209.8 - 1219.8 \AA\ and the continuum subtracted to measure the total line flux, as is described in Section~\ref{sec:firstresults}. The ratio of the fluxes $\eta$ for each \delLAE{} cutoff are shown in Figure~\ref{fig:line_ratios}. Note that we are not enforcing the minimum area requirement of 40 Mpc$^2$, which is an additional requirement to define a protocluster region in the surface density maps of \cite{ramakrishnan_odin_2025}.


For almost all \delLAE{} cutoff values at both redshift ranges, Figure~\ref{fig:line_ratios} shows that $\eta > 1$, indicating that galaxies in above the \delLAE{} cutoff value have more luminous \lya{} lines on average than galaxies in regions of lower overdensity. Above \delLAE{} $\approx 2.3$, $\eta$ is comparable between the N419 and N501 galaxies. Both datasets have a maximum value of $\eta \approx$ 1.3, quite close to the value of 1.35 which was reported by \cite{dey_spectroscopic_2016} as the \lya{} line luminosity enhancement in a protocluster region at $z \approx 3.786$. Note that as higher \delLAE{} cutoff values are approached, the ratios become noisier as there are fewer galaxies meeting the new ``protocluster" criterion and thus the associated boostrapped error bars are larger. In the N501 sample, there are 451 galaxies above \delLAE{} = 2 and 12 above \delLAE{} = 6, while in N419 there are 532 and 7 above \delLAE{} = 2 and 6, respectively.

The luminosity function of LAEs, and thus their SFRs, evolve with redshift, although the exact nature of this relationship is not fully understood \citep{ouchi_observations_2020, nagaraj_odin_2025}. Although LAEs are broadly thought to be highly star-forming, they experience peaks and troughs of star formation at different points throughout their lifetimes \citep{rosani_bright_2020, firestone_odin_2025}. Comparisons of SED fits suggest that LAEs at $z \approx$ 3.12 do not necessarily evolve to remain LAEs at $z \approx$ 2.1, but rather that LAEs at different redshifts are totally different sets of galaxies \citep{acquaviva_curious_2012}. The difference in the in vs. out of protocluster line profiles between the N419 and N501 galaxies suggests that our sample of LAEs may be at a different point in their star formation histories between $z \approx$ 2.45 and 3.12, but further investigation is required to reach a definitive conclusion.

LAEs exhibit a wide range of line profiles, including features such as secondary blue peaks or troughs with different amounts of separation, velocity offsets, varying levels of asymmetry, and more \citep{vitte_muse_2025}. Aside from the potential \lya{} line luminosity and EW differences, Figure \ref{fig:501_stacks} shows that the line profile shape in the ODIN data barely varies with environmental density -- there is little change in the slope of the line or any other features on either side of the \lya{} wavelength. This suggests that the factors causing other variations in the \lya{} line profile shape, such as AGN outflows and resonant scattering of \lya{} photons, occur at different spatial scales than what we can probe using the NB data from ODIN. This evidence for the separation of line effects at different distance scales bodes well for the potential use of LAEs as cosmological probes, suggesting that LAEs in the same field at the same redshift likely have very similar intrinsic line profiles.


\section{Conclusion}
\label{sec:conclusion}
We present an investigation of the relationship between environmental density and the \lya{} line profile in DESI spectra of ODIN LAEs selected with NB photometry and z $\approx$ 2.45 and 3.12. By comparing stacked \lya{} line profiles of LAEs inside and outside of protocluster regions identified by \cite{ramakrishnan_odin_2025}, we can compare the shape and strength of the \lya{} line at varying levels of environmental density. Our conclusions are as follows:
\begin{itemize}
    \item LAEs observed at $z \approx$ 3.12 in the COSMOS field have $\sim$15\% higher \lya{} line luminosities in protocluster regions than outside of protocluster regions. 
    \item This potential relationship between environmental density and \lya{} line luminosity indicates elevated star formation in protocluster galaxies.
    \item This luminosity enhancement within protocluster regions is not observed as strongly in the lower redshift $z \approx$ 2.45 LAEs. This could indicate a redshift evolution of galaxy-environment interactions, but is also potentially reproduced by imposing higher surface density cutoffs.
    \item Aside from the line luminosity and EW, the shape of the \lya{} remains largely similar across varying environmental densities, with no change in presence of a blue peak, line asymmetry, or velocity offset.
\end{itemize}

Overall, the DESI spectra of LAEs selected with NB photometry from the ODIN survey suggest a relationship between local environmental density and the \lya{} line profile, but further investigation with larger datasets would provide additional insight into the nature of this relationship.

\section*{Data availability}
The data for all figures in this paper are available at \doi{10.5281/zenodo.17651863}.

\begin{acknowledgments}
ASMU was supported by a National Science Foundation Graduate Research Fellowship and would like to thank Daniel Eisenstein, Haruki Ebina, Matthew Hayes, Alberto Saldana-Lopez, Mark Dickinson, Christina Williams, and Ashley Ortiz for helpful conversations.

A.D.’s research activities are supported by the
NSF NOIRLab, which is managed by the Association of
Universities for Research in Astronomy (AURA) under a
cooperative agreement with the National Science Foundation.

 NF acknowledges support from NSF grant AST-2206222 and NSF Graduate Research Fellowship Program under Grant No. DGE-2233066.

LG gratefully acknowledges financial support from ANID - MILENIO - NCN2024\_112, ANID BASAL project FB210003, FONDECYT regular project number 1230591.

HSH acknowledges the support of the National Research Foundation of
Korea (NRF) grant funded by the Korea government (MSIT),
NRF-2021R1A2C1094577, and Hyunsong Educational \& Cultural Foundation.

H.S. was supported by the National Research Foundation of Korea (NRF) grant funded by the Korea government (MSIT) (No. RS-2025-25442707).

This work is supported by the National Science Foundation under Cooperative Agreement PHY-2019786 (The NSF AI Institute for Artificial Intelligence and Fundamental Interactions, \href{http://iaifi.org/}{http://iaifi.org/}).

This research used resources of the National Energy Research Scientific Computing Center (NERSC), a Department of Energy Office of Science User Facility.

This material is based upon work supported by the U.S. Department of Energy (DOE), Office of Science, Office of High-Energy Physics, under Contract No. DE–AC02–05CH11231, and by the National Energy Research Scientific Computing Center, a DOE Office of Science User Facility under the same contract. Additional support for DESI was provided by the U.S. National Science Foundation (NSF), Division of Astronomical Sciences under Contract No. AST-0950945 to the NSF’s National Optical-Infrared Astronomy Research Laboratory; the Science and Technology Facilities Council of the United Kingdom; the Gordon and Betty Moore Foundation; the Heising-Simons Foundation; the French Alternative Energies and Atomic Energy Commission (CEA); the National Council of Humanities, Science and Technology of Mexico (CONAHCYT); the Ministry of Science, Innovation and Universities of Spain (MICIU/AEI/10.13039/501100011033), and by the DESI Member Institutions: \url{https://www.desi.lbl.gov/collaborating-institutions}. Any opinions, findings, and conclusions or recommendations expressed in this material are those of the author(s) and do not necessarily reflect the views of the U. S. National Science Foundation, the U. S. Department of Energy, or any of the listed funding agencies.

The authors are honored to be permitted to conduct scientific research on I'oligam Du'ag (Kitt Peak), a mountain with particular significance to the Tohono O’odham Nation.

\end{acknowledgments}

\begin{contribution}
A. S. M. Uzsoy was responsible for the analysis and paper writing, A. Dey \& A. Raichoor were responsible for the spectroscopic observations and reduction of spectroscopic data, A. Dey \& D. Finkbeiner were primary advisors for the analysis, V. Ramakrishnan \& K-S. Lee provided the LAE surface density maps used in the analysis, K-S. Lee and E. Gawiser are PIs of the ODIN collaboration, N. M. Firestone, L. Guaita, H. S. Hwang, and H. Song are builders of the ODIN collaboration and all remaining authors are builders of the DESI collaboration.


\end{contribution}

%
\facilities{Mayall (DESI), Blanco (DECam)}

\software{Julia \citep{bezanson2017julia}, {\tt zELDA} \citep{zelda}}

\appendix
\section{N419 stacks \& medians separated by field}
\label{appendix:separate_Stacks}

\begin{figure*}[ht!]
\epsscale{1.15}
\plottwo{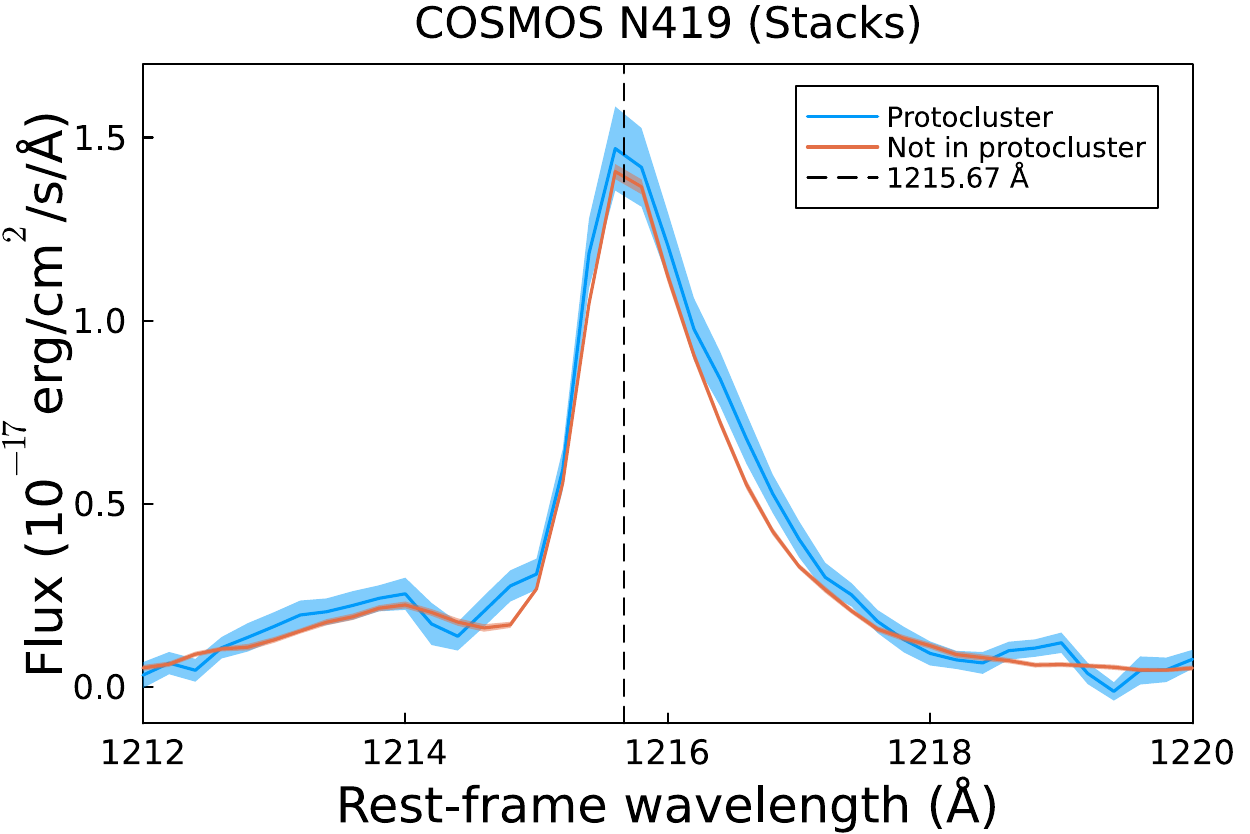}{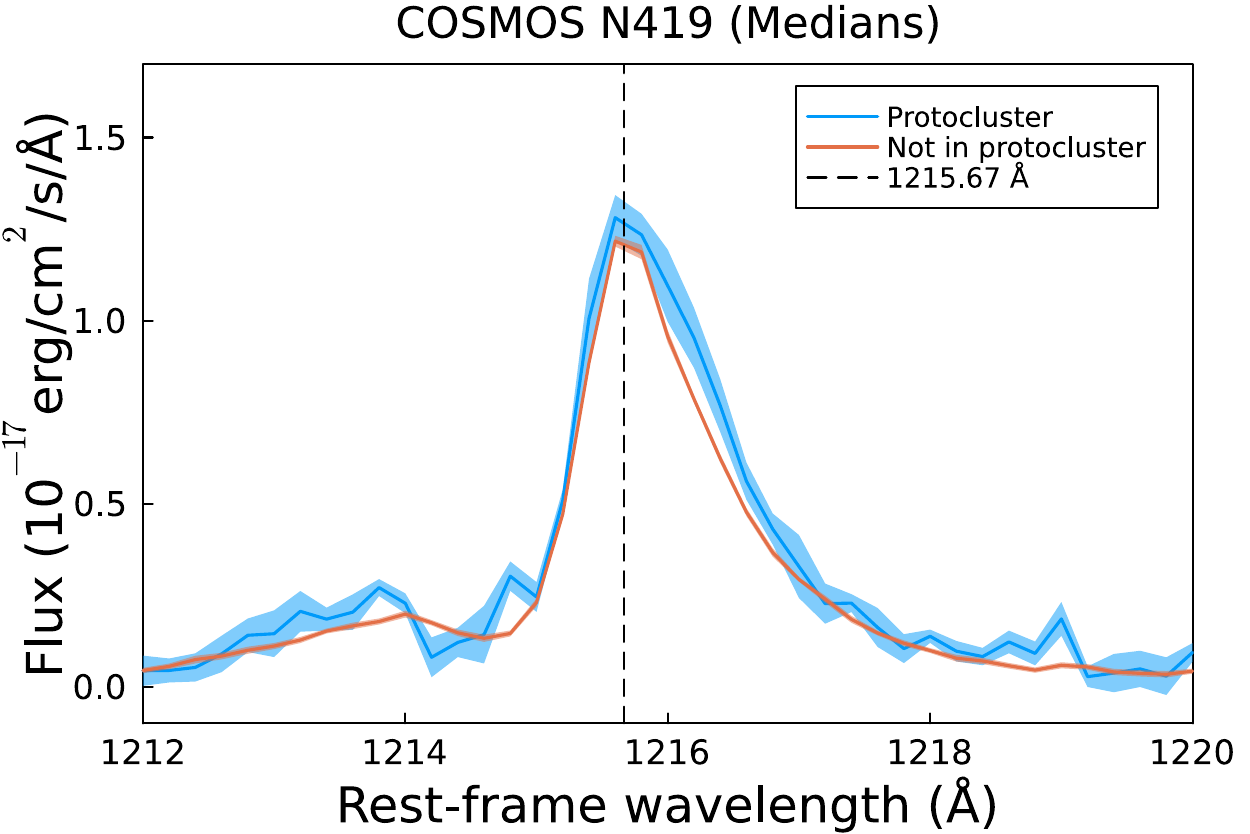}
\caption{Inverse-variance-weighted mean stacks (left) and median spectra (right) of the Ly$\alpha{}$ profile for ODIN LAEs in the N419 band, inside (blue) and outside (orange) of protocluster regions in the COSMOS field.}
\label{fig:419_cosmos_proto_stacks}
\end{figure*}

\begin{figure*}[ht!]
\epsscale{1.15}
\plottwo{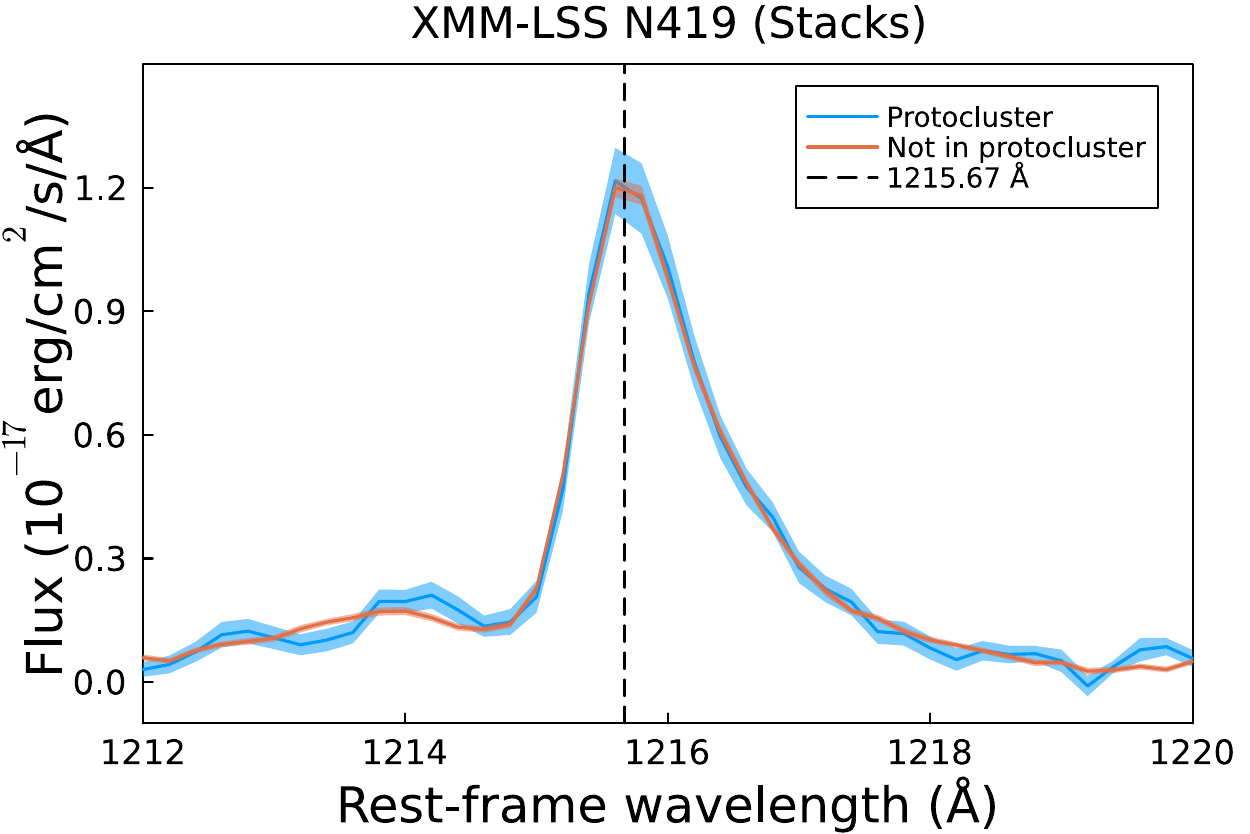}{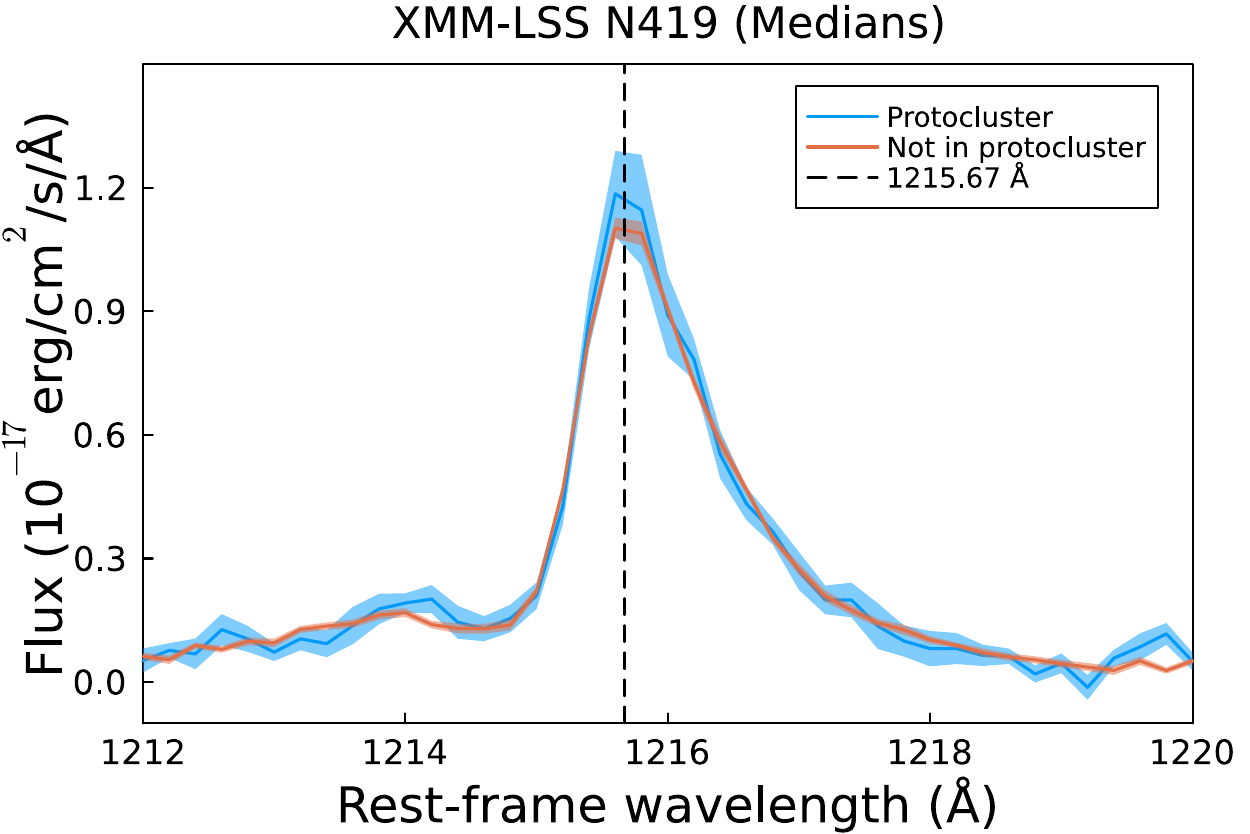}
\caption{Inverse-variance-weighted mean stacks (left) and median spectra (right) of the Ly$\alpha{}$ profile for ODIN LAEs in the N419 band, inside (blue) and outside (orange) of protocluster regions in the XMM-LSS field.}
\label{fig:419_xmm_proto_stacks}
\end{figure*}

Figure~\ref{fig:combined_419_proto_stacks} shows the stacked and median spectra inside and outside of protocluster regions in galaxies selected with the N419 band ($z \approx 2.45$) combined across the COSMOS and XMM-LSS fields. Figures~\ref{fig:419_cosmos_proto_stacks} and \ref{fig:419_xmm_proto_stacks} show the in vs. out of protoclusters stacks and medians for N419 galaxies selected in the COSMOS and XMM-LSS fields, respectively. Both datasets separately and combined (see Figure~\ref{fig:combined_419_proto_stacks}) do not show a significant difference in \lya{} line flux between LAEs inside and outside of protocluster regions.

\section{Selection of double-peaked line profiles}
\label{appendix:double_peaks}

\begin{figure}[ht!]
\plotone{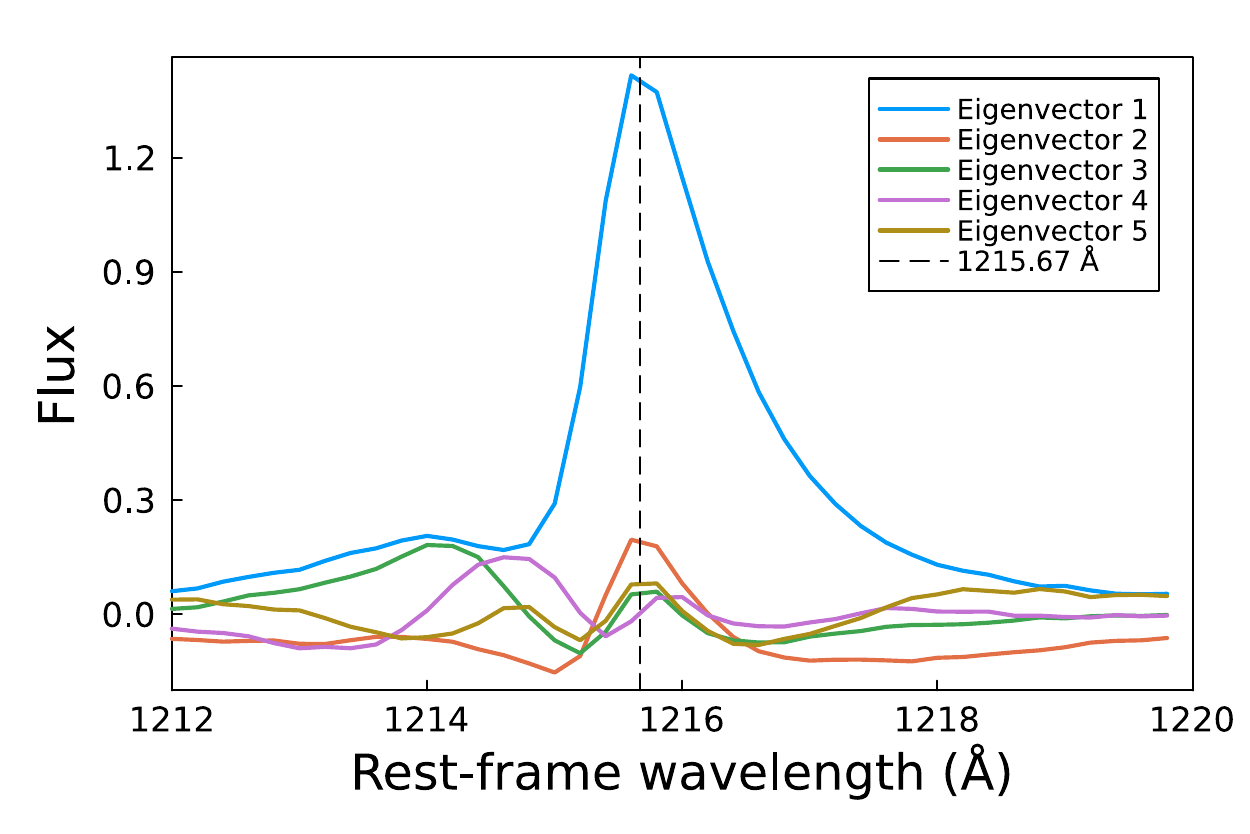}
\caption{Top five eigenvectors of the data-driven covariance matrix of ODIN LAEs shifted to rest-frame, scaled by the square root of their respective eigenvalues. }
\label{fig:eigenvecs}
\end{figure}

To create a data-driven model of the \lya{} line, we first create a data-driven covariance matrix of the LAE spectra. We use the method outlined in Section 3.1.2 of \cite{uzsoy_bayesian_2025}, but with no normalization by a sky polynomial, linearly-spaced wavelength bins, and with all spectra shifted to rest-frame. The variance of the flux values of each spectrum was calculated, and the top 5\% were excluded in the creation of the covariance matrix to ensure only high SNR spectra were used. Figure~\ref{fig:eigenvecs} shows the top five eigenvectors of this data-driven covariance matrix. Each individual galaxy's \lya{} line can then be expressed as a set of coefficients on these eigenvectors, where coefficients of the third and fifth eigenvectors are mainly driven by the presence of a secondary blue peak. 

\begin{figure}[ht!]
\plotone{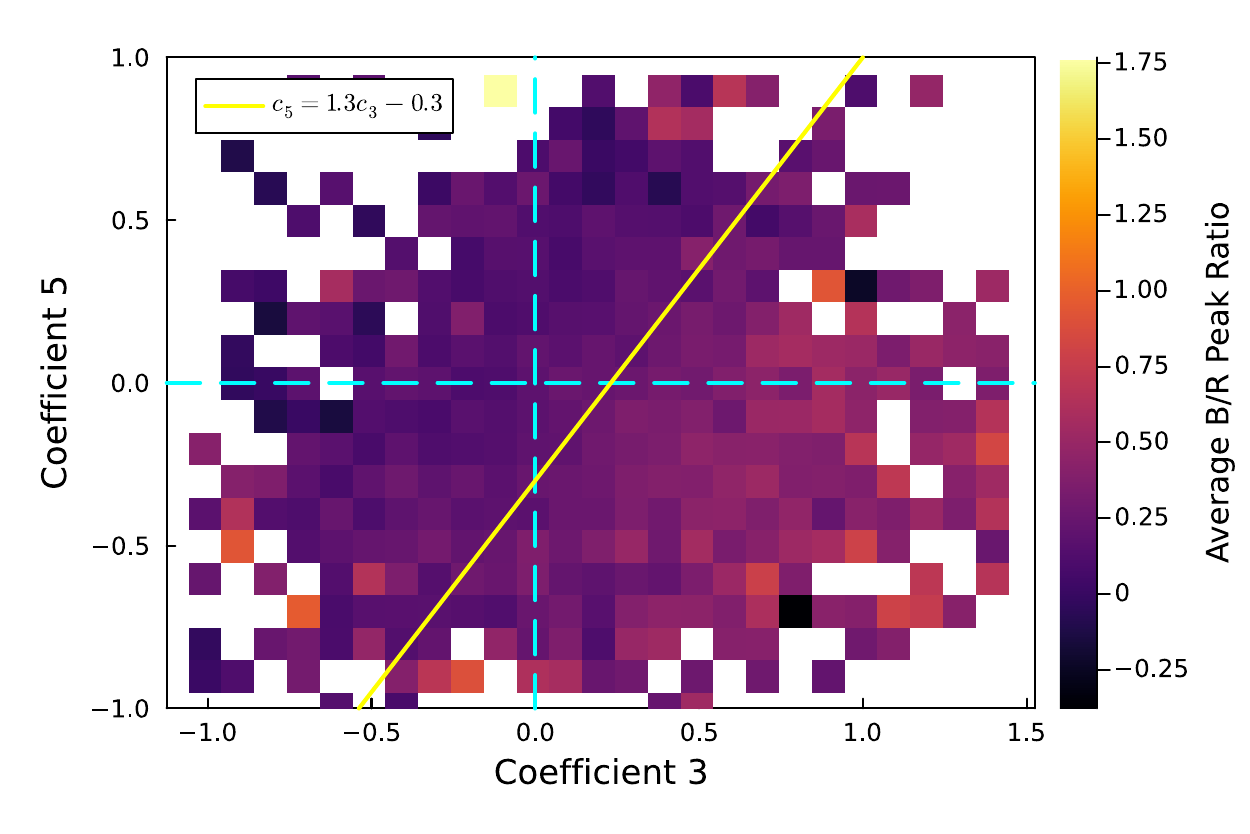}
\caption{ODIN LAEs in both the N419 and N501 bands, binned in the coefficients of the third and fifth eigenvectors, with the colorbar denoting the average blue to red peak ratio in each bin. Yellow line denotes the dividing line to identify double-peaked LAEs, and cyan dotted lines denote zero for both coefficients. }
\label{fig:c3c5}
\end{figure}

\begin{figure}[ht!]
\plotone{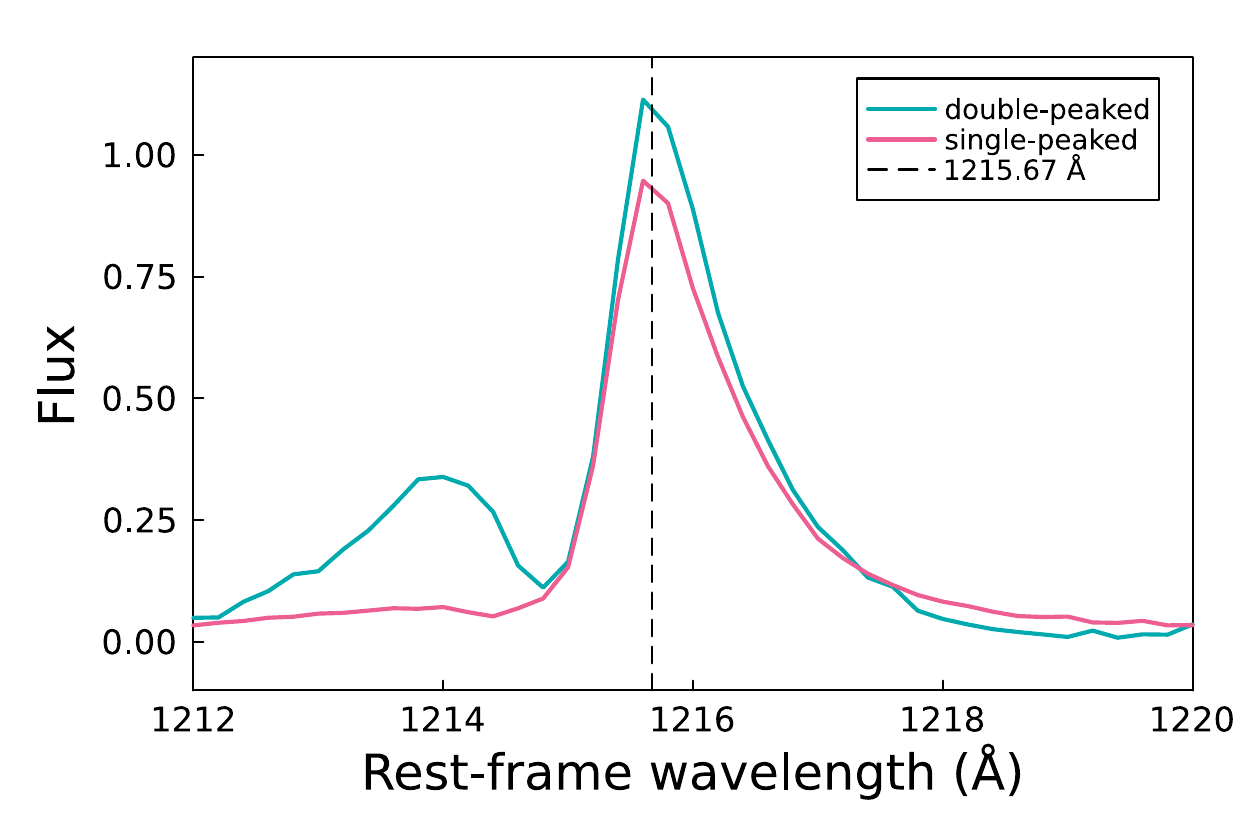}
\caption{Inverse-variance-weighted mean stacks of the Ly$\alpha{}$ profile for N501 ODIN LAEs that have been identified as single vs. double peaked.}
\label{fig:single_double_stacks}
\end{figure}

As a point of comparison, we also fit this dataset with {\tt zELDA} \citep{zelda}, a machine learning tool trained on radiative transfer simulations to identify the \lya{} line and measure its properties. Figure~\ref{fig:c3c5} shows the average blue-to-red peak ratio fit by {\tt zELDA} in bins of our third and fifth eigenvector coefficients. We draw a thresholding line in this space and identify any galaxies with $c_5 < 1.3 c_3 - 0.3$ as double-peaked. Figure~\ref{fig:single_double_stacks} shows stacks of galaxies (selected with either the N419 and N501 band) that were identified as single- vs. double-peaked using this criterion, which was then used in identifying the double-peaked N501 galaxies shown in Figure~\ref{fig:double_stacks}.

\section{Complex A}
\label{appendix:complexA}

\begin{figure}[ht!]
\epsscale{1.15}
\plottwo{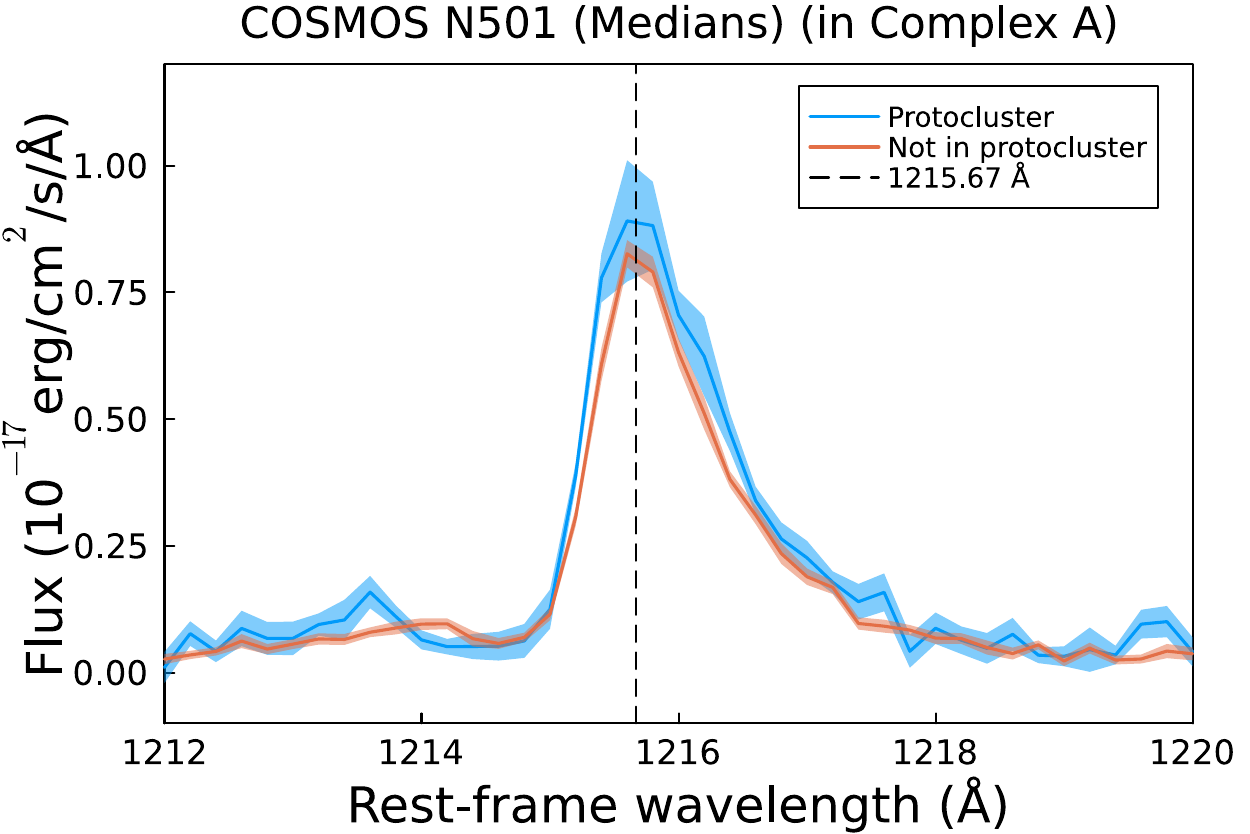}{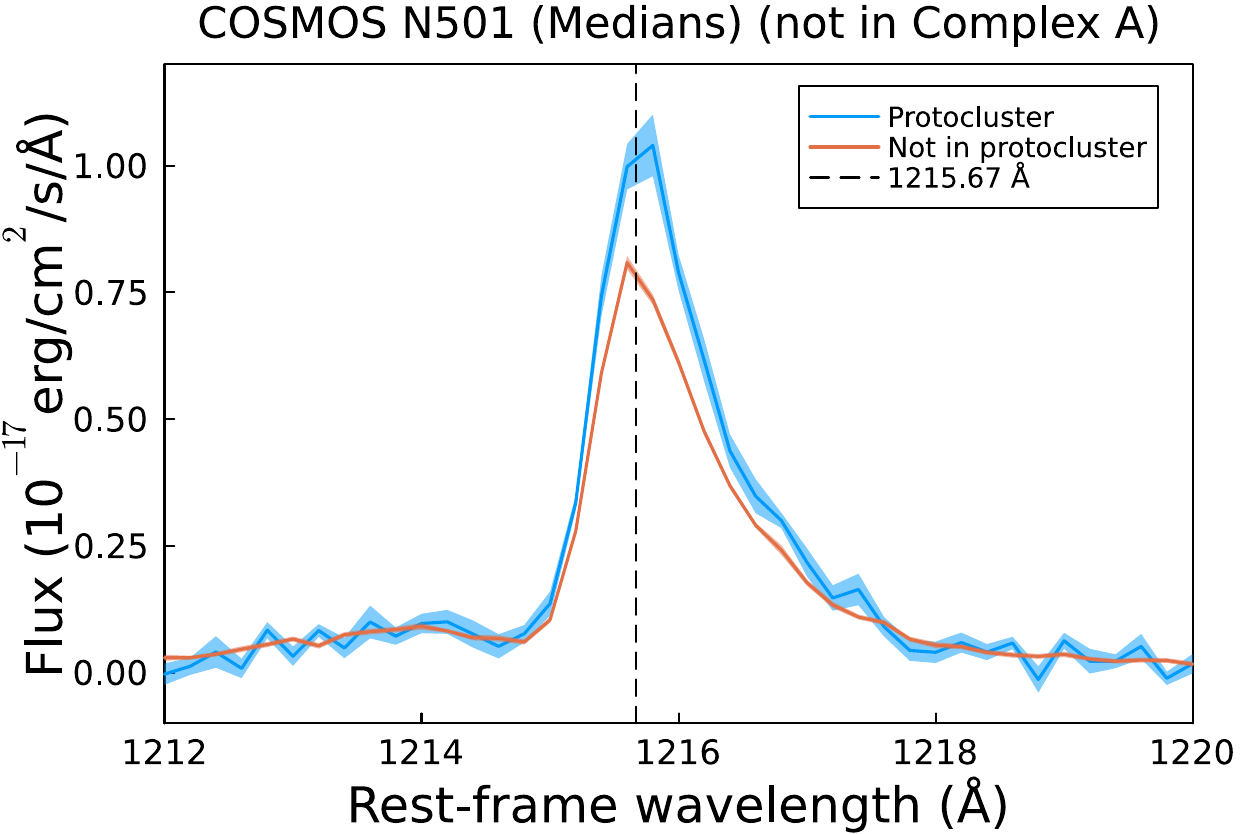}
\caption{Median spectra of the Ly$\alpha{}$ profile for N501 ODIN LAEs inside (left) and outside (right) of Complex A, inside (blue) and outside (orange) of protocluster regions in the COSMOS field.}
\label{fig:complexA}
\end{figure}

In the COSMOS field in the N501 NB filter, there exists a region of extreme overdensity, known as Complex A \citep{ramakrishnan_odin_2023, ramakrishnan_odin_2025, ramakrishnan_complexA}. We distinguish galaxies in this region by selecting all galaxies with RA between 150.2 and 151.4 degrees, and DEC between 2.2 and 3.2 degrees \citep{ramakrishnan+24}. Using this criteria, we identify 333 galaxies with DESI spectra in Complex A, 46 of which are in protoclusters. To ensure that our results in Figure~\ref{fig:501_proto_stacks} are not solely driven by the presence of this ``proto-supercluster", we conduct the same analysis for galaxies inside and outside of Complex A, seen in Figure~\ref{fig:complexA}. The same effect of in-protocluster galaxies having significantly higher \lya{} luminosity can be seen even when Complex A is excluded from the analysis.

\bibliography{sample7}{}
\bibliographystyle{aasjournalv7}



\end{document}